\documentclass[a4paper,10pt]{article}

\RequirePackage{latexsym}
\RequirePackage[english]{babel}
\RequirePackage{amssymb}  
\RequirePackage{amsbsy}   
\RequirePackage{amsmath}  
\usepackage[T1]{fontenc}
\usepackage[dvips]{graphicx}

\newcommand{\ud}{\mathrm{d}}
\newcommand{\udd}{\frac{\mathrm{d}}{\mathrm{d}t}}
\newcommand{\uddn}{\frac{1}{\gz}\,\udd}
\newcommand{\ls}{speed of light }
\newcommand{\gzz}{c(t)\,\ud t + t\,\ud c}

\newcommand{\gz}{(c+t\dot{c})}
\newcommand{\gznp}{c+t\dot{c}}

\newcommand{\beq}{\begin{equation}}
\newcommand{\eeq}{\end{equation}}
\newcommand{\bspl}{\begin{split}}
\newcommand{\espl}{\end{split}}
\newcommand{\dota}{\dot{a}}
\newcommand{\gra}{\grave{a}}
\newcommand{\axz}{a(x^0)}
\newcommand{\ek}{{k_{\!E}}_{\scriptstyle 0}}

\author{Corrado Appignani%
\thanks{E-mail: \textsf{appignani@bo.infn.it}}\\[3mm] 
\emph{Dipartimento di Fisica --
Universit\`a di Bologna}\\[1mm] \emph{Via Irnerio 46 -- I-40126 --
Bologna -- Italy}}
\title{A geometrically-induced varying speed of
light (VSL) and the accelerating universe}

\begin{document}
\maketitle
\noindent

\begin{tabbing}
\> Acceleration of the universe, Curvature scalar, 
Varying fine-structure constant\\[1mm]
\end{tabbing}

\begin{abstract}

We tackle the problem of the accelerating universe by
reconsidering the most general form of the metric when the
speed of light is allowed to evolve with time in a
homogeneous and isotropic universe. A new varying speed of
light (VSL) model naturally emerges. We find it unconvenient
to perform a general coordinate transformation to obtain the
usual constant $g_{tt}$ as such an operation would mask the
relation existing between the evolution of the scale factor
and the frequency shifts. In the model proposed the
expansion rate and the acceleration of the universe turn out
to be apparent effects induced by the evolution of the speed
of light. The model is beneficial in that no sort of exotic
(and so far unobserved) fluids, not even a cosmological
constant, are needed for our solutions to be compatible with
observations. Only dust and radiation are put into the
energy-momentum tensor and these are found to be sufficient
to reach the critical density for the model and therefore to
obtain a spatially-flat universe. The field equations for
the model are derived and solved. Among others, one
fashinating possibility is that of an eternally bouncing
universe. The relation with the varying-$\alpha_\textsc{em}$
results is discussed.

\end{abstract}

\pagebreak


\section{Introduction} 
\label{sec:intro}

\begin{quote}
\emph{%
``It conflicts with one's scientific understanding to conceive a
thing which acts but cannot be acted upon.''}

A.~Einstein on the concept of absolute space.
\end{quote}

Many authors have recently proposed cosmological models
where a variation with cosmological time of the universal
constants, in particular of the speed of light, is
hypothesized as a viable alternative to the widely accepted
inflationary scheme in order to solve the classical
problems of cosmology (see \cite{Magueijo} and references
therein for a comprehensive review of the topic). 

Attempts in this direction started about ten years ago
with Moffat \cite{Moffat} who considered a sudden
change in the \ls and discussed the cosmological
implication of such a phase transition. The same kind
of variation was used a few years later by Albrecht
and Magueijo to propose their $c$-varying model that,
under certain conditions, presents no cosmological
problems. A model for the universe with $c$ evolving
according to a power law was proposed in the same year
by Barrow \cite{Barrow}, who also looked for solutions
to the field equations and set the constraints under
which his solutions solve the classical problems.  

Other authors also opted for a continuously varying
$c(t)$, see for instance \cite{Belin}, where
Belinch\'on and Alfonso-Faus introduced the Planck's
constant in the field equations via the
Stefan-Boltzmann law, discussed the gauge invariance
for the Schr\"odinger, Klein-Gordon-Fock and Maxwell's
equations and showed that their model does not present
the so-called Planck's problem. See also the
``revised'' version \cite{belin2}. In other works (see
e.g.~Bel in \cite{Bel}), the \ls directly acquires the
role of the reciprocal of the scale factor, with the
metric divided by $c^2$, and the dimensionality of a
length squared  preserved through multiplication by
an appropriate constant. 

Another interesting way to introduce a varying \ls is the
bimetric theory of gravity proposed by Clayton and Moffat
\cite{mocla} where the issue of diffeomorphism invariance is
elegantly approached by assuming the presence of two
different metrics which implies a different speed for
gravitons and photons. This model is capable to solve the
classical cosmological problems as easily as inflation does
but, contrary to inflation, it does not need fine tunings or
unnaturally flat potentials\footnote{For a review about the
most important problems with the inflationary paradigm see,
for instance, \cite{Lue}.}. More recently \cite{mocla2},
Clayton and Moffat have shown that their bimetric theory
predicts a scale-invariant spectrum for cosmological
perturbations which is compatible with CMB data but
distinguishable from the inflationary prediction.

Despite some evidences for a varying fine-structure constant
$\alpha_{\textsc{em}}$ \cite{Webb1,Webb2,Webb3,Mur1,Mur2}
(contested for instance in \cite{noalpha} and rebutted in
\cite{sialpha}), which at the time
being represent the only observative evidence for the
variation with redshift (and therefore with time) of
(at least) one ``universal constant'', there are some
authors who contest the very meaning of stating that
the \ls might be not constant (see e.g.\ in
\cite{Duff}). The most popular objection seems to be
that, after all, it all boils down to the choice of a
system of units through which one can see a
varying-$\alpha$ theory as if $h$ or $e$ were time
dependent instead of $c$, or together with $c$, or
finally in any possible combination of the three (see
also \cite{ellis} for a reactionary view on varying-$c$
theories). This objection deserves consideration
even if, in the general case, the appropriate
redefinition of units has to be continuous in, say, the
cosmic time. 

After some 15 years from their ``discovery'', the
discussions about the meaning of VSL theories is going on. A
recent opposers' viewpoint may be found in \cite{EllisList}
where a sort of general check list aimed to decide if a VSL
model merits consideration at all is proposed. This list
however does not seem to take into account the different
ways and purposes to introduce a VSL mechanism, and the true
intention seems to show that introducing a VSL is equivalent
to upset the very basics of modern physics, which is
certainly not the case in general and surely is not true for
the model presented in this paper. 

The arguments presented in \cite{EllisList} have been
quickly replied in \cite{MagMoff} where a particularly
illuminating example can be found (see section II). It
relates to acceleration of gravity, the once-believed
``constant'' $g$, and makes clear how the arguments put
forth by the opposers of VSL theories could have well
applied at the time of Newton against any proposals that $g$
were not constant. This wonderful example should make clear
to anyone how such choices really depend on the knowledge
of physics at a given time. Saying that $c$ is constant
because so we have decided (so we have defined units) is not
different from insisting that $g$ is constant because so
Galileo had decided. As Galileo was wrong because at his
time he could not measure $g$ at a large distance from the
Earth surface, so we may well be wrong because in no way we
can know \emph{now} what the speed of photons was $10^9$
years ago (see the next section for what we exactly mean
with ``the speed of photons has varied''). We do not know if
the speed of light has varied or not, but it is sure that it
is a possibility that worths be investigated.

As said, the reported evolution of $\alpha_{\textsc{em}}$ may
also be attributed to a varaition of $e$ or $\hbar$.  An
example of a time-dependent $\alpha_{\textsc{em}}$ described
by a varying-$e$ theory, we address the works by Bekenstein
\cite{beke1,beke2} and their generalization by Barrow,
Magueijo and Sandvik \cite{bsbm} (the BSBM model). Sometimes
the variation of $\alpha_{\textsc{em}}$ has been considered
together with that of $G$. Time-dependent $G$ theories date
back to the well-known works by Dirac \cite{dirac} and by
Brans and Dicke \cite{bd}. More recently Barrow and Magueijo
have studied the behaviour of Brans-Dicke cosmologies with a
varying speed of light \cite{baema}. See also
\cite{bouncing} for a parallelism between Brans-Dicke
cosmologies and varying speed of light theories.

While, for the reason given above, a variable fine-structure
constant -- whose variation has been recently contested, see
\cite{chand} and references therein --  cannot be taken as
an evidence for a variation of the speed of light with
cosmological time, it is certainly incorrect to consider a
variable \ls meaningless or tautological. The whys, clearly
explained in \cite{Magueijo} (chapter two), are easily
summarized as follows. Since $\alpha_{\textsc{em}}$ is a
dimensionless quantity, no matter which system of units one
chooses to adopt, if $\alpha_{\textsc{em}}$ evolves with
time, it will do so whatever the units employed. As a
consequence, if $\alpha_{\textsc{em}}=e^2/c \hbar$ depends
on cosmic time then at least one of the universal
``constants'' $c$, $h$ or $e$ must depend on cosmic time as
well. Suppose now to have a varying-$\alpha_{\textsc{em}}$
theory with, say, a time-dependent $c$, and that such a
theory has a simple dynamical structure. It is then
certainly true that, by means of an appropriate units
transformation, one can always rewrite that
varying-$\alpha_{\textsc{em}}$ theory so as to have a
constant $c$ and, say, an evolving $e$, but by doing so one
may end up with an unnecessarily contrived formalism which
was much more elegant and simple in the varying-$c$ scheme.
Besides, given that $\alpha_{\textsc{em}}$ is dimensionless
and therefore cannot be made constant by means of any
transformation of units whatsoever, it is clear that in the
new system of units $e$ and/or $h$ would be time dependent,
and it is not clear why the variation of $e$ and $h$ should
be regarded as less problematic than the variation of $c$.

I wish to remark that this ``simplicity'' argument is not
(as some might think) too weak to justify a varying-$c$
cosmology. Let me remind that even the most primitive
definitions adopted in physics are justified by similar
arguments, the most outstanding example being perhaps the
difinition of time (see \cite{grav} par 1.5 page 23). We
define time by means of regular and periodic phenomena not
because it could not be done otherwise, but because we want
a unit of time which is the same now and tomorrow, a feature
which facilitates a comparison between physical phenomena
which are distant in time and guarantees that the equations
of dynamics look simple.

It is also to be considered that the experimental tests
proposed in \cite{isitcore} show how varying-$c$ and
varying-$e$ models may behave differently and could
then in principle be distinguished if the experimental
precisions were to reach a sufficiently high level.
Proposed tests span from the weak equivalence principle
to the dark matter needed to fit quasar data to
earth-based gravitational redshift experiments
(Pound-Rebka-Snyder). For all these tests, Magueijo,
Barrow and Sandvik \cite{isitcore} have shown that
varying-$c$ and varying-$e$ models predict different
when not opposite results. Finally, when a more
complete and fundamental theory of physical phenomena
will be available, we could eventually discover that
one or more of what we now believe to be fundamental
constants may in fact be derived quantities or just one
of the factors combining to form a more fundamental
constant. This, of course, would spoil that quantity of
the role of ``fundamental constant''. 

To the best of our knowledge, assuming that the
evidence for a variation of $\alpha_{\textsc{em}}$ will
receive independent confirmations, it is presently
impossible to predict which of the three constant
$\alpha_{\textsc{em}}$ is made of is fundamental or
not. We can only make guesses based on simplicity
arguments. This principle of simplicity is the one that
we will follow in the present work. It will be seen
that one cannot rewrite this model in terms of a 
constant $c(t)$ without paying the price of
complicating the dynamics (e.g.~by working with a
continuously changing system of units) and making
comparison with observations unnecessarily indirect and
complicated. 

So far, works on VSL theories have mainly focused on showing
that these theories are a viable alternative to the
inflationary scheme, especially when dealing with the
solution to the classical problems of cosmology. In this
paper we present a VSL model that provides an interesting
solution to the puzzle of the acceleration of the universal
expansion discovered some years ago \cite{riess} (for a more
recent paper see \cite{sn}).

Instead of introducing exotic and so-far-unobserved forms of
dark matter-energy, we introduce only radiation or dust as
sources in the energy-momentum tensor. We show that a
universe filled with such ``normal'' content (and where the
speed of light is varying) appears to accelerate even if the
actual state of motion is not that of an accelerated
expansion. Na\"ively, this is due to the fact that
instruments measure frequencies but the dynamics of the
scale factor is ``represented'' by the evolution of
wavelengths (see appendix A). \emph{Because of the relation
$\lambda \nu = c$, if the speed of light varies with
cosmological time, the inverse proportionality between
wavelengths and frequencies is lost and apparent effects
arise}.

The paper is organized as follows. In section
\ref{sec:postulates} we derive the general metric to be used
under the approach proposed in the paper. In section
\ref{sec:appscalef} we introduce the concept of apparent
scale factor and discuss how it is linked to the classical
scale factor. In section \ref{sec:tensors} we give the
expressions for the main tensorial quantities deriving from
the metric proposed. Section \ref{sec:appacc} contains the
phenomenological core of the paper: we show that the
acceleration of the universe may be an apparent effect and
derive under which circumstances such an effect takes place.
In section \ref{sec:fieldeqs} we solve the field equations
in the case of spatial flatness and show that a universe
composed of dust and/or radiation only actually meets the
condition derived in section \ref{sec:appacc} under which an
apparent acceleration is observed. In section
\ref{sec:conschecks} we perform some consistency checks to
verify that the model proposed is both internally consistent
and consistent with the relevant observative results. The
summary is given in section \ref{sec:conclu}.


\section{Postulates of the model and natural emergence 
of a VSL}
\label{sec:postulates}

Most of the non-bimetric VSL models have been developed
postulating that the speed of light has varied with
cosmological time, either in a sharp phase transition
or continuously. Such variation is assumed to affect
the local conservation equation (the effect is
``special relativistic''), but \emph{not} the field
equations \cite{AM}. The dependence of the \ls $c$ on
cosmological time is simply accounted for by making the
direct substitution $c \rightarrow c(t)$ wherever
appropriate.

On the contrary, in our approach the form of the metric is 
affected by the evolution of $c(t)$. As we have found that
this position is considered questionable by many who are
accostumed with standard cosmology, we dedicate this section
to discuss the issue. As is well known, general relativity
exhibits an invariance under general coordinate
transformations. This implies, for instance, that we could
in principle operate an appropriate transformation of the
time variable so as to obtain a metric where $c$ is
constant. The advantage of operating such a transformation
would clearly be the reduction of the degrees of freedom
from two to just one, the scale factor $a(t)$, and the usual
FLRW metric would then be recovered. However obvious the
naturalness of such an operation could appear, performing it
is not at all compulsory, again because of general
covariance. The choice is one of convenience and is dictated
by the advantages and disadvantages implied by it. The
question should be: does such an operation imply a global
advantage or a global disadvantage? The answer depends of
course on what the purposes of one's work are. We will find
that, as far as comparison between the results of the model
with observative evidences is concerned, operating the
mentioned transformation would create more interpretative
problems than the computational problems it would save us
from. Therefore we find it simply not convenient to change
the coordinate set, just as we do not use a meter stick
whose length oscillates, increasing, for instance, during the
day and decreasing during the night: while perfectly
possible, comparison of lengths of other objects and its use
in general would become an awkward task.

Let us see examine the details of the problem. We start as
usual with a four dimensional manifold endowed with some
arbitrary metric and label a generic set of coordinates with
$x^0$, $x^1$, $x^2$ and $x^3$, all having the dimension of a
length.  The last three are chosen to be some ``spatial''
coordinates of convenience, for instance the spherical
coordinates, while the $x^0$ coordinate must represent time
in some way. In view of the cosmological application we next
consider the restrictions imposed on it by assuming the
validity of the cosmological principle. In the case of
spatial flatness\footnote{The extension to the spatially
curved case is straightforward.}, the most general line
element compatible with the cosmological principle is:
\beq
{\ud s}^2 = b^2(x^0) (\ud x^0)^2 - 
{\axz}^2{\vec{\ud r}}^2 
\eeq
where $b(x^0)$ and $\axz$ are some function of $x^0$. Note
that we have chosen the $x^0$ direction in such a way to
make the $g_{ti}$ elements vanish\footnote{This is not
different from what one does in standard cosmology upon
choosing synchronous time as a time coordinate of
convenience.}. Here we may set $b(x^0)=1$ with no loss of
generality as such an operation amounts to a redefinition of
the meter, which cannot affect physical results (what we can
measure is the ratio  ${\axz}/b(x^0)$). We then obtain:
\beq
{\ud s}^2 = (\ud x^0)^2 - {\axz}^2{\vec{\ud r}}^2
\label{lineel0}
\eeq
where we have not bothered to rename the variable $x^0$ and
the function $\axz$. The choice made implies that
\emph{no further simplifications can arise from changing
units after this stage,} otherwise one should put
${b(x^0)}^2$ or some other appropriate function back into
the metric.

The next step is to introduce time in the metric, both for
sake of visualization (after all we do think in terms of
time) and to make a comparison with the observation, in
particular, with the redshifts or frequency shifts. What is
the most convenient choice in our case, that is, in the case
where one wants to consider the possibility of a
time-dependent speed of light? One constraint is given by
the fact that experience shows that physics is
special-relativistic at a local level so that  $x^0 = tc$
must hold locally. Here $c$ is the locally\footnote{With
``locally'' we are meaning here at a given $x^0$, that is,
on a given hypersurface of simultaneity. } measured speed of
light and $t$ is the proper time along fluid worldlines
(this time variable is global if space-time is flat).  The
general transformation which is compatible with both this
local limit and the cosmological principle is therefore
$x^0(t) = tc(t)$ where $c(t)$ reduces to the locally
measured speed of light $c$. We remark that here we are
substituting Lorentz invariance with local Lorentz
invariance, see \cite{magLLI}.

From $x^0(t) = tc(t)$ it follows that
\beq
\ud x^0 = (c(t)+t\dot{c}(t))\,\ud t
\label{eq:dexz}
\eeq
so that, in terms of the familiar coordinates, 
$g_{tt}$ becomes:
\beq
g_{tt} = (c+t\dot{c})^2
\eeq
where we have dropped the time dependence stipulating
that, from now on, $c$ is always to be considered 
time-dependent unless otherwise stated.
This implies that, under the restrictions imposed by
the cosmological principle, the most general line
element to be used in this varying-$c$ model is
\beq
{\ud s}^2 = (c+t\dot{c})^2 {\ud t}^2 - 
a^2 {\vec{\ud r}}^2
\label{eq:metricf}
\eeq
where, we repeat, spatial flatness has been assumed.

At this point one would be tempted to redefine the time
variable so as to obtain a constant speed of light: $c(t) =
c_0$. However, contrary to what happens in standard
cosmology, performing such an operation happens to be not
convenient in our case, and we dedicate the remainder of
the paragraph to see why.

We first discuss the importance of considering the local
limit. By construction, the $t$ variable used in the
previous expression is the time of special relativity and,
therefore, the time ticked by clocks at rest in a given
reference frame. It follows that, in a cosmological setting,
this time variable can be used as  ``the proper time in the
comoving frame'' that is, the time measured by clocks at
rest with respect to the reference frame comoving with the
cosmological fluid. If we assume that the relevant emission
and absorption properties have not changed with time, then
the use of such time variable guarantees that emission and
absorption frequencies are numerically the same now and at
any time. This property is important because it is precisely 
the requirement a time variable must fulfill for the observed 
redshifts to be completely ascribed to the cosmological 
expansion.

We remark that this property remains true if and only if we
use, up to a constant factor, the time variable given above.
Indeed any coordinate change $t \rightarrow t'$ aimed to
reduce  $g_{t't'}$ to a constant would have to be
time-dependent, $t'=t'(t)$ (any space dependence is
forbidden by the cosmological principle), and this would
imply, among other things, that all emission frequencies,
\emph{if expressed in terms of the new $t'$ variable}, could
be different at different times $t'$, i.e.~moving from one
spacelike hypersurface to another. On the contrary, by using
the time $t$, we are sure that we are using \emph{physical}
frequencies. It is evident that the crucial drawback of some
transformation of the time variable would be that we could
not anymore compare properties of radiation emitted in the
past with what is known from earth-laboratory physics in a
straightforward way. In particular any estimation of
redshifts would necessitate a distance-dependent correction
before comparison to earth-laboratory physics could be
made. 

To see this point more clearly, let us attempt the
conversion to an appropriate coordinate set to obtain a
constant speed of light. We could for instance transform
the time variable $t$ in the following way: 
\beq
(c+\dot{c}t_{\text{old}})^2 \ud t_{\text{old}}^2 
\,\longrightarrow\, c_0^2 \ud t_{\text{new}}^2. 
\eeq 

In this new reference frame the speed of light is constant
but what are the consequences of such a transformation?
Probably the most evident is that while any given physical
phenomenon (say, an atomic transition) which in terms of the
$t_{\text{old}}$ variable  lasted $\bar{t}_{\text{old}}$
seconds at some epoch of the past will, by construction,
also last $\bar{t}_{\text{old}}$ seconds today, the same
cannot be true in the $t_{\text{new}}$ frame or in terms of
any other time variable different form $t_{\text{old}}$ (up
to a constant factor). The unpleasant consequences of the
choice of some given $t_{\text{new}}$ are evident: the
measured frequency shift experienced by any given radiation
would have to be ascribed to two different contributions,
one being the usual cosmological expansion effect and the
other being the continuously changing emission/absorption
frequency of the related atomic transition (due, as some put
it, to the ``evolving time unit'') to which we compare the
radiation to get the redshift. Only the first effect could
be related to the dynamics of the scale factor and, as the
measured redshift -- which is measured by antennae which, of
course, measure frequencies and not wavelengths -- would
carry the imprints of both effects, we would have to
disentangle these two effects. Depending on the exact
expression for $t_{\text{new}} =
t_{\text{new}}(t_{\text{old}})$, the comparison with the 
observations would require a distance-dependent correction
which could turn out to be an arbitrarily complicated and
useless task as, \emph{because of general covariance, the
results concerning the physical observables would in the end
be the same}. 

One conclude then that while in the $t_{\text{old}}$ frame
the comparison between the theory and the observations is,
by construction, straightforward, in the $t_{\text{new}}$
frame it could turn out to be unpredictably complicated.
Since, because of general covariance, physical results, and
in particular the evolution obtained for the physical
observables, would turn out to be the same, there is no good
reason to introduce a new, unconvenient time variable. This
is why we will stick to the most simple choice and will keep
working with the $t_{\text{old}}$ variable.

This discussion should have made clear the simplicity
argument we have been referring to: we use the varying speed
of light frame and the related time unit because \emph{it is
in this frame and only in this frame that the observed
wavelength shift is purely cosmological}. We close by noting
that this remark allows to define what we do mean with a
``varying speed of light'' in a somewhat explicit form.
Consider a physical phenomenon by which radiation of a given
frequency is emitted. Think of the same phenomenon taking
place at some time in the past and today. We have adopted
the unit/frame by which the frequencies in the two cases are
the same. Observe, for the two cases separately, the length
a light ray can travel in, say, a unit of the time variable
we have adopted. \emph{We will say that the speed of light
has varied if that length is different for the two
situtations.}


\section{The apparent scale factor}
\label{sec:appscalef}

The most enlightening way to study the consequences of
the modification to the $tt$ coefficient of the metric
induced by the variable speed of light and obtained in
the last section is probably that of finding a relation
to link the evolution of the scale factor $a(t)$ (which
represents the actual dynamics of the universe) with
another variable, which conveniently represents the
evolution of the scale factor as seen by an observer
who is not taking the variation of the speed of light
into account (henceforth referred to as the
``constant-$c$ observer'').  We will denote such
variable with $A(t)$ and will refer to it as the
\emph{apparent scale factor}. 

As a first step toward this goal we have to re-express the
evolution of $\axz$ in terms of that of $a(t)$. In fact the
former will be seen to play the role of a bridge which
connects the actual and the apparent scale factors.  From
the evolution of $\axz$ we will then be able to derive the
evolution of the apparent scale factor. 

As already said, $A(t)$ represents the evolution of the
scale factor as seen by constant-$c$ observers. In practice,
this means that $A(t)$ is  what is usually obtained when
estimating the evolution of $a(t)$ from observations. As
will be made clear later, it is the confusion between $a(t)$
and $A(t)$ which gives rise to the apparent effect of the
acceleration in the expansion of the universe: $A(t)$
``accelerates'' while $a(t)$ (the universe) does not.

A preliminary remark will be helpful in understanding what
follows. We recall that, whatever the method we may choose
to perform any measurement on a cosmological scale, we have
to use distant objects or, better, the radiation we receive
from them. This radiation reaches us by travelling through
space-time, along the worldline which separates the two
events: (A) radiation emitted by the object, (B) radiation
received by our instruments. Photons from these objects or
from the LSS (Last Scattering Surface) have travelled their
long way through space-time to reach our satellites and, in
doing so they have experienced, event after event, the
dynamics of the gravitational field, both the spatial and
time metric coefficients continuously varying along their
paths. The direct consequence of this fact is that, in our
model, the effects we measure cannot be ascribed to the
variation $a(t)$ alone. In particular, in a continuosly
varying-$c$ theory, it is incorrect to identify the shift in
the frequencies of a photon with the evolution of the scale
factor $a(t)$. 

As anticipated, a way to tackle this problem is to account
for such additional dynamical effects by providing the
relation existing between the evolution of the scale factor
and that of the \emph{apparent} scale factor.  To accomplish
this task, the first step is to calculate $\ud
\sqrt{g_{rr}}/\ud x^0$ in terms of $\dot{a}(t)$. What we
obtain is
\begin{equation}
\frac{\ud \sqrt{g_{rr}}}{\ud x^0} = 
\frac{\ud}{\gzz}\; a(t) =  \frac{\ud}{\gz \, \ud t} \;
a = \frac{\dot{a}}{\gz}. 
\label{eq:axzdiat}
\end{equation}

This result will provide us with a connection between the
observed and the actual dynamics of our universe. Let us
denote this quantity with $\gra=\gra(t)$ just to have a
concise notation\footnote{We choose not to use the prime
$'$ to avoid confusion with the established notation which
denotes the derivative with respect to conformal time.}
for the last term in equation (\ref{eq:axzdiat}). We may 
then rewrite the last equation as
\begin{equation}
\gra = \frac{\dot{a}}{\gz}
\label{eq:msdef}
\end{equation}

In what follows we will show that this variable works
like a sort of bridge between $\dot{a}(t)$, the
dynamics of the universe, and $\dot{A}(t)$, which is
what is actually obtained when estimating $\dot{a}(t)$
from measured frequency shifts of the radiation coming from
distant objects. Before we can give the exact relations
which link these two variables and their derivatives,
we need to calculate the basic tensor quantities for
the present model. This is done in the next section.


\section{Basic tensor quantities}
\label{sec:tensors}

The new metric in spherical coordinates for a FLRW universe with
a time-dependent speed of light is:
\begin{equation}
g_{\mu \nu} = \
\left( \begin{matrix} 
      \big[ c(t) + t\dot{c}(t) \big] ^2 & 0 & 0 & 0 \\ 
      0 & \displaystyle - \frac{{a^2(t)}}{1 - kr^2}  & 0 & 0 \\ 
      0 & 0 & - r^2 a^2(t) & 0 \\ 
      0 & 0 & 0 & -r^2 a^2(t) \sin^2 (\theta) 
\end{matrix} \right)
\end{equation}
which yields the following connection coefficients with a
contravariant $t$ index:
\beq
   \Gamma^{t}_{\phantom{t} \beta \gamma} = 
   \left(
   \begin{matrix}
    \displaystyle
    \frac{\partial_t[c + t\dot{c}]}{c + t\dot{c}} & 0 & 0 & 0 \\
    0 & \displaystyle \frac{a\dot{c}}{\left( 1 - kr^2 \right) 
    {\left(c + t\dot{c} \right) }^2} & 0 & 0  \\
    0 & 0 & \displaystyle \frac{r^2 a \dota}
    {{\left(c + t\dot{c} \right)}^2} & 0   \\
    0 & 0 & 0 & \displaystyle \frac{r^2 a \dota \,{\sin^2 (\theta )}}
     {{\left(c + t\dot{c} \right) }^2}
   \end{matrix}
   \right)
\eeq
   
The connection coefficients with a spatial 
contravariant index are the same as those for the FLRW
metric. In fact $g_{tt}$ depends on $t$ only and no
further changes have been made in the usual FLRW
metric, thus the ``spatial'' $\Gamma$'s are left
untouched.

\noindent
The Ricci tensor is diagonal and its components are:
\begin{align}
& R_{tt} = 
     \,\frac{-3\,\gz}{a}\, \udd \left[ 
     \frac{\dota}{c+t\dot{c}} \right]\\[5mm]
& R_{rr} = \frac{1}{(1-kr^2)}\,
    \left[ 2k+ \frac{1}{a\,\gz}\, \udd 
    \left(\frac{a\dota^2}{c+t\dot{c}} \right) \right] \\[5mm]
& R_{\theta \theta} = r^2 \, (1-kr^2) \, R_{rr} \\[5mm]
& R_{\phi \phi} = r^2 \, \sin^2(\theta) \, (1-kr^2)\, R_{rr} \,
  = \, \sin^2(\theta) \, R_{\theta \theta}
\end{align}

\noindent
The components of the Einstein tensor are:
\begin{align}
& G^t_{~t} = \frac{3}{a^2} \cdot 
\left[ k+ \frac{\dota^2}{\gz^2} \right] \\[5mm]
\begin{split}
& G^r_{~r} = G^{\theta}_{~\theta} = G^{\phi}_{~\phi} = \\[3mm]
& = \frac{1}{a^2} \left[k+ \uddn \left( \frac{a \dota}{\gznp} \right)
+ \frac{a}{\gz} \, \udd  \left( \frac
{\dota}{\gznp} \right) \right] =  \\[3mm]
& = \frac{1}{a^2} \left[ k+ \frac{1}{a \dota}\, \udd 
\left[ \left( \frac{a \dota}{\gznp} \right)^2 \right] -
\frac{1}{a^2}\, \left( \frac{a \dota}{\gznp} \right)^2 
\right]=\\[3mm]
& = \frac{1}{a^2} \left[k+ \frac{1}{a}\, \udd 
\left( \frac{\dota}{\gz^2} \right) +
\frac{\dota}{a \gz^2}\, \udd \left[ \log(a \dota)
\right] \right]= \\[3mm]
& = \frac{1}{a^2} \left[k+ \frac{2a}{\gz}\, \udd 
\left( \frac{\dota}{\gznp} \right) +
\frac{\dota^2}{\gz^2} \right]= \\[3mm]
& = \frac{1}{a^2} \left[k+ \frac{1}{\dota}\, \udd 
\left[ \left( \frac{a \dota}{\gznp} \right)^2 \right]
\right]~.
\end{split} \end{align}
where we have given several different expressions for
$G^r_{~r} = G^{\theta}_{~\theta} = G^{\phi}_{~\phi}$
as each of them turns out to be useful depending on
the situation. In some of these expressions
non-vanishing denominator(s)/argument(s) are assumed
so their use is subject to a preliminary check of
such condition(s).

\noindent
Finally, the curvature scalar is conveniently expressed as:
\beq
R \, =\,-\,\frac{6}{a^2} \cdot \bigg[
  k + \uddn \! \left( \frac{a \dota}{\gz} \right)
\bigg]~.
\label{eq:cs} 
\eeq


\section{Apparent acceleration in a flat universe}
\label{sec:appacc}

Here we establish the existence of apparent effects that, in
our model, are measured by constant-$c$ observers. We
discuss the relation between the actual and the apparent
evolution of these scale factors. We will show that such a
universe appears to be in an accelerated expansion
($\dot{A},\ddot{A}>0$), when its actual motion is that of an
accelerated contraction, i.e., when the scale factor $a(t)$
is a decreasing function of $t$.\footnote{As we will soon
show, this is not in contradiction with the Hubble law.}

As already remarked, any measurement made by observing some
kind of radiation coming from another event in space-time is
a test of the evolution of both the spatial and the time
components of the metric.  Therefore, what we effectively
estimate is the apparent motion of the universe as desumed
from frequency measurements performed by our instruments. Of
course all of the existing estimations have been performed
without taking into account a possible variation of $c(t)$
with cosmological time. It follows that any direct
comparison between observations and the theory are
intrinsically flawed if $c(t)$ is time-dependent. In fact, a
direct comparison inevitably leads to a wrong guess for the
behaviour of $a(t)$. More precisely, it leads to a confusion
between $a(t)$ and $A(t)$.

What we set out to prove now is that a homogeneous and
isotropic universe with a varying speed of light and
filled with ``normal'' content only (that is, without
exotic matter or a cosmological constant) \emph{may}
appear to be in an accelerated expanding motion, this
apparent effect being caused by a variation of the
speed of light.

We provide two ways to show what we have just claimed. The
first way is rather simple and makes use of the relation
existing between the wavelength and the frequency of a given
wave which, in the case of an electromagnetic wave
propagating in vacuum is: $\lambda \nu = c$. It consists in
calculating the apparent and the real evolution of the
wavelength starting from a given measurement of the
frequency of the same wave and considering the variation of
$c$ with time. It is clear that this method is well founded
only if we are making a comparison between what is recorded
when observing radiation coming from a distant object and
what is known about atomic physics and atomic spectra. The
second method is more general and applies to any possible
case. The two methods give the same results.

\subsection{Motivations for space-time flatness}

Before looking into the first proof we need an intermediate 
result. In fact, the
starting point of both proofs is to require that the
curvature scalar vanishes. In this paper we limit ourselves
to the study of a spatially flat universe
($k=0$)\footnote{The spatially curved case will be discussed
in a paper under preparation.} so that the requirement
becomes
\beq
R(k\!=\!0)\!=\!0.
\eeq
This assumption may appear controversial so we will
dedicate a few lines to discuss it in some detail.

First we remark that, while the vanishing of the Ricci
scalar is \emph{assumed}, we can easily \emph{argument}
that the Ricci scalar, whatever its value, should remain
constant. In fact the gravitational interaction, in its
geometrical representation, is described by the
dynamics of a four-dimensional manifold which, in a sense, ``mimic''
a gravitational field. The curvature of this manifold
is expressed by the Riemann tensor. By contracting 
this tensor two times, one obtains the Ricci scalar
which, therefore, is the scalar quantity that best
represents the curvature of the manifold \cite{gauss}. 

In a cosmological setting, for a homogeneous and isotropic
universe, the form of the metric is that of
Robertson-Walker, i.e., that of a maximally symmetric space
so that, in the FLRW as well as in our model (which can be
seen as a VSL extension of the FLRW model) the Ricci scalar
$R$ can, in principle, depend on time $t$ only, and not on
the spatial variables: $R=R(t)$.

Next consider that General Relativity is based on the
Einstenian conception that gravitation in nothing but the
curvature of space-time, and that this curvature can be
generated by mass-energy-momentum and only by it. Given
that, by its very definition, the universe cannot interact
and exchange mass-energy-momentum with anything else, it
immediately follows that the average space-time curvature of
the universe should not be allowed to change with time. As
the space-time curvature can be expressed, for instance, by
the Ricci scalar -- which because of the cosmological
principle cannot depend on the spatial variables -- we
arrive at the conclusion that the Ricci scalar must remain
constant in time.

We stress that all this discussion \emph{applies to a
homogeneous and isotropic cosmological setting only}. In
fact, in different situations, the possibility of exchanging
energy-momentum with an ``outside'' forbids to follow the
above line of reasoning and therefore the Ricci scalar
cannot be constrained if not in a cosmological ambit.

The fact that $R$ must remain constant does not however mean
that it must vanish. The value zero has been \emph{chosen}
essentially for two reasons. First, somewhat
philosophically, we may say that it seems to be a natural
choice: an initial condition must be chosen and deciding
that the universe ``has come into existence'' with no prior
space-time curvature appears to be, in a sense, the most
natural one. Second, \emph{more formally}, if we want a
classical universe to emerge from its quantum epoch (and
from a possible inflationary stage) in a radiation-dominated
state, then we must fix $R=R_\textsc{rad}$ as a sort of
matching condition. As during this stage the trace of the
energy-momentum tensor vanishes $T \equiv T^\mu_\mu =0$ it
follows that $R=0$ for any model based on Einstein field
equations. When this is combined with the previous remarks,
we are led to fix $R=0$ once and for all.

We will soon see that the fact that some forms of matter,
e.g.~dust, do not satisfy $T \equiv T^{\mu}_{\ \mu}=0$
(which is implied by $R=0$) does not represent a problem for
our model. Indeed the variation of $c$ will be seen to
induce additional terms in the components of the
energy-momentum tensor. These additional terms will in turn
induce a solution to $T \equiv T^\mu_{\ \mu} = 0$ which is
different from $\epsilon=3p$ or from the trivial one,
$\epsilon = 0$ (here $\epsilon$ is the energy density), so
that \emph{fixing $R=0$ does not constrain the equation of
state in the model we are proposing}.

We digress for a moment to note that the assumption $R=0$,
when made in the standard FLRW cosmology, enforces a
radiation-like equation of state (or that of an empty
universe). In fact, $R_{\textsc {flrw}} = 0$ implies $a(t)
\propto t^{1/2}$ and this shows that a flat FLRW space-time
seems to be compatible only with a permanently
radiation-dominated universe. The ``creation of a net amount
of curvature'' taking place locally at every point of the
spacelike hypersurface of simultaneity in the standard FLRW
cosmology at the transition from the radiation to the matter
dominated epoch (during which the Ricci scalar decreases as
$1/t^2$) is hard to explain when one adopts the viewpoint
outlined above.  Since gravity couples to any kind of
energy-momentum, and since the total amount of
energy-momentum must obviously be conserved during the
transition from the radiation-dominated ot the matter
dominated-phase, it follows from the contracted field
equations that the mean space-time
curvature represented by the Ricci scalar must be conserved
as well. As already said, this problem does not arise in
this VSL model, where one can impose the vanishing of the
Ricci scalar and still retain the possibility to have any
kind of equation of state that satisfies the positive energy
condition.

\subsection{Consequences of space-time flatness}

The assumption of space-time and spatial flatness, when
combined, turn out to be very useful. From equation
(\ref{eq:cs}), keeping only the last term which does not
contain $k$, we get
\beq
R \,= \ \frac{-6}{a^2\,\gz} \; 
 \udd \left[ \frac{a \dota}{\gz} \right] \, = \, 0
\label{eq:Rcost}
\eeq
which provide us with a fundamental relation between the
scale factor and the \ls\!:
\beq 
\frac{a \dota}{\gz} \, = \, \text{\sc{constant}}.
\label{eq:const1} 
\eeq
The general solution for the scale factor to equation 
\eqref{eq:const1} is
\beq 
a(t) \, = \, a_0
\sqrt{\frac{tc}{t_0c_0}}
\label{eq:gensol}
\eeq
and it is immediate to realize that the only
\emph{elementary} function fulfilling such requirement is a
power law. This could have also been seen by noting that, for
the left hand side to be constant, that is, independent from
$t$, the first requirement is that both terms in the
denominator must have the same dependence on $t$. This can
be so if and only if $c(t)$ evolves according to a power
law. This forces the numerator to be also a power law
function. So, both the scale factor and the speed of light
must evolve according to a power law if the evolution law
for the universal expansion has to be simple.

Imposing for $a(t)$ and $c(t)$ the following forms:
\begin{align}
& a(t) \, = \, \frac {a_0}{t_0^N} 
\, t^N \label{eq:adit}\\[1mm]
& c(t) \, = \, \frac {c_0}{t_0^M} 
\, t^M \label{eq:cdit}
\end{align}
with $N,M \in \mathbb{R}$, we get
\beq 
 \left(\frac {a_0}{t_0^N} \right)^2
 \left(\frac {t_0^M}{c_0} \right)
 \frac{N\,t^{2N-1}}{(1+M)\,t^M} \, = 
 \, \text{\sc{constant}}
\eeq 
which immediately yields
\begin{align}
 &M \, = \, 2N-1 \label{eq:mandn}\\
 &\textsc{constant} \, = \, \frac{a^2_0}{2t_0c_0}
 \end{align}
so that equation (\ref{eq:const1}) becomes
\beq 
\frac{a \dota}{\gz} \, = \, \frac{a^2_0}{2t_0c_0}
\label{eq:const} 
\eeq

We remark that using $\sqrt{g_{tt}}=c(t)$ instead of $\gz$
does not allow to obtain the previous relations so that the
whole of the results that follows cannot be extended to
other continuosly-varying speed of light theories. It is
also not possible to draw the conclusion that $c(t)$ and
$a(t)$, must be power law functions.

\subsection{An intuitive argument}
\label{subsec:anintarg}
Keeping the last result at hand, let us now suppose that
we have measured the properties of the radiation
coming from a set of distant objects. In particular
let us assume that we have measured the frequency of a
given emission or absorption line, compared it with
what we already know from atomic physics about that
process, and have finally deduced a dependence of that
frequency on cosmological time which can be written as
$\nu \propto t^{-S}$ with $S \in \mathbb{R}$. As a
deduction, if thinking in terms of a constant speed of
light, we get that $A(t) \propto \lambda \propto t^S$
(let us remember that $A(t)$ is the apparent scale
factor, apparent in the sense that it is what
is observed by a constant-$c$ observer).

Now let us think in terms of a varying speed of light.
We have just seen that, if the scale factor behaves as
$t^N$, then the speed of light is proportional to
$t^{2N-1}$, see (\ref{eq:cdit}) and \eqref{eq:mandn}. From
$\lambda \nu = c$, we then obtain that frequencies are
proportional to $t^{N-1}$. 

As our measurements gave $\nu \propto t^{-S}$, we get
that $N=1-S$ and not $S$ is the correct exponent in 
the power law evolution of the wavelength and thus for
the scale factor. In other words the scale factor
evolves like $t^{1-S}$ but appears to be evolving 
like $t^S$:
\beq
\text{\sc{if}} \quad a(t) \propto t^{1-S} \qquad
\text{\sc{then}} \quad A(t) \propto t^S.
\label{eq:ascafa}
\eeq
As 
\beq
\ddot{A}(t) > 0 \quad \longleftrightarrow \quad S<0 \ \lor \
S>1, 
\eeq
we deduce that a positive acceleration is detected
when $N\!>\!1$ or $N\!<0$.  In particular the second
possibility tells us that a contracting universe with
a varying speed of light manifests itself as a
universe that is undergoing an accelerated expansion,
and this is exactly what we set out to show.

The other solution, yielding $N\!>\!1$, is to be discarded
for several reasons. For instance, it implies $S\!<\!0$,
that is, the observation of a contracting universe which is
in contradiction with the experience. Also, it implies an
increasing speed of light, which is really unwanted as the
speed of light must decrease for the horizon problem to be
elegantly solved by a VSL mechanism. But most importantly we
will not consider the possibility $N\!>\!1$ since, as we
shall see later, it will be formally excluded upon solving
the field equations. Indeed, it will be shown that each
solution to the field equations satisfies $N \in [-1/2,1/2]$
which clearly excludes $N\!>\!1$.

We remark that the result given above concerning the
apparent nature of the acceleration has been achieved with
no assumption about the content of the universe. \emph{No
exotic matter and/or cosmological constant have been
introduced. The effect is simply induced by the variation of
the speed of light.}

\subsection{A general proof}

The argument exposed in the preceding section is not
rigorous and may not apply in some cases.  There is however
a more general way to show how and when an observed
acceleratation turns out to be an apparent effect. We make
use of (\ref{eq:msdef}) to calculate the evolution for
$\gra$ in terms of the cosmological time:
\beq
\gra \equiv \frac{\dot{a}}{\gz} = 
\frac{a_0}{2c_0} \, t_0^{N-1}t^{-N}
\eeq
The evolution of $\gra$ is a way to represent the dynamics
of the manifold and, in a sense, it provides a bridge
between the real and the apparent evolution of the scale
factor expressed in terms of cosmological time. In fact, an
observer who does not take into account the variation of
$c(t)$ would perceive an evolution for the apparent scale
factor given by
\beq
\dot{A} = c \mspace{1mu} \gra = 
\frac{a_0}{2t_0^N}\,t^{N-1}.
\eeq
The crucial point to note is that $\dot{A}$ is always
positive, irrespective of the value of $N$. This is a very
important property of this model: \emph{no matter whether
the universe is expanding or contracting, the observed
(apparent) motion will always be found to be that of an
expansion}.

This means that we may well be living in a varying-$c$
universe whose spacelike hypersurfaces are undergoing a
contracting motion, but such a universe would anyhow show
just the same Hubble law a standard FLRW universe shows.
Therefore, and we stress this very important point, the fact
that $N\!<0$ \emph{is not in contradiction with the
experimental evidence that we observe redshifts and not
blueshifts}.

The result for the apparent second derivative of the
scale factor is even more interesting. We make use of
the previous result for the apparent first derivative
of the scale factor to derive  the apparent
acceleration:
\beq 
\ddot{A} \, = \, c \,
\grave{\left( \mspace{-1mu} c \, \grave{\!\!\!a\ } \right)}
\, =  \, 
\frac{N-1}{N} \, \frac{a_0}{4t_0^N} \, t^{N-2}
\eeq

As this seems to be a source of confusion, we remark that
$\ddot{A}$ is not the time derivative of $\dot{A}$ just as
$\dot{A}$ is not the time derivative of $A$. In words,
$\dot{A}$ is  the apparent time-derivative of the scale
factor, \emph{not} the time-derivative of the apparent scale
factor: $\dot{A} = (\dot{a})_\textsc{app} \neq
\frac{\ud}{\ud t}(a_\textsc{app})$.

The previous equation shows that an acceleration for
the expansion of the universe is measured if and only
if $\ N\!<\!0 \ \lor \ N\!>\!1$ :
\beq
\text{\sc{positive acceleration observed}} \quad
\Longleftrightarrow \quad
N\!<\!0 \ \lor \: N\!>\!1. 
\label{eq:acc} 
\eeq
Note that this is exactly the same result we have
obtained with the first method. We made no use of the
properties of a wave propagating through space-time. It
is a simple computation of how modification of
distances induced by the cosmological expansion are
seen by an observer who does not take into account the
variation of the speed of light.

The crucial point to note is that $\dot{A}$ can be positive
even when $N$ is negative. It is precisely this result that
allows $N$ to be negative, thus causing a positive
acceleration, without the need for $\dot{A}$ to be negative
as well, which of course would mean to \emph{observe} a
contraction, in contradiction with the experimental evidence
that we observe redshifts and not blueshifts.  This is
evident by looking at the summary for the evolution of the
(apparent) derivatives of the three variables $a(t)$, $\axz$
and $A(t)$ which is given in Table \ref{tab:evolution}.

\begin{table}[!h]
\begin{center}
\begin{tabular}{|r||l|l|l|l|}
\hline
& \multicolumn{2}{|c|}{(Apparent) 1$^{\textsc{st}}$ 
derivative}
& \multicolumn{2}{|c|}{(Apparent) 2$^{\textsc{nd}}$ 
derivative}\\[1mm] 
\hline
\hline
$a(t)$ & $N\,t^{N-1}$ & $>0$ ~for~ $N\!>\!0$ &
$N(N-1)\,t^{N-2}$ & $>0$ ~for~ $N\!\!<\!0 \, \lor
N\!>\!1$\\[1mm]
\hline
$\axz$ & $\frac{1}{2}\,t^{-N}$ & $>0 ~~ \forall N$ 
& $-\frac{1}{4} \, t^{-3N}$ & $ <0 ~~ \forall N$\\[1mm]
\hline
$A(t)$  & $\frac{1}{2}\,t^{N-1}$ & $>0 ~~ \forall N$ &
$\frac{N-1}{4N}\,t^{N-2}$ & $>0$ ~for~ $N\!\!<\!0 \, 
\lor N\!>\!1$\\[1mm]
\hline
\end{tabular}
\end{center}
\caption{Time evolution for the three ``scale factors'' in
terms of time. Note that $\dot{a}\!<\! 0$ if $N\!\!<\!0$ but
$\dot{A}\!>\! 0 ~ \forall N$. We stress that $\ddot{A}$  is
not the time derivative of $\dot{A}$ just as $\dot{A}$  is
not the time derivative of $A$. In words, $\dot{A}$ is  the
apparent time-derivative of the scale factor, \emph{not} the
time-derivative of the apparent scale factor.}
\label{tab:evolution}
\end{table}

Hidden behind the above line of reasoning is the assumption
that there is no variation of the emission properties with
cosmological time. Indeed, this feature derives from the
choice of the time variable and is therefore automatically
guaranteed in our model.\footnote{Note that the same is
instead \emph{assumed} in standard cosmology.} If that was
not the case, we could not make any comparison between
atomic spectra obtained on earth laboratories, and
emission/absorption lines of distant objects. This issue has
already been discussed in section \ref{sec:postulates} but a
further discussion is perhaps at order, given that the
characteristics of the emitted radiation depend on the fine
structure constant $\alpha_{\textsc{em}}$ and that, as
mentioned in the introduction, some recent papers (the last
one being \cite{Mur2}) have reported a very tiny relative
variation of this constant. We postpone this discussion to
section \ref{sec:conschecks} where the consistency of the
model with observations and its predictions are addressed.


\section{Field equations and their solutions}
\label{sec:fieldeqs}

We have shown that a universe that appears to be
undergoing an accelerated expansion actually is,
according to the present model, in a state of
contracting motion. Yet we still have to prove that
such a behaviour is a solution to the field equations
of the model and discuss what the content of the
universe has to be in order to obtain it. Our next task
will then be that of solving the field equations.
However, one runs into difficulties right from the
start. In fact in this model an additional problem
arises, that of expressing the energy-momentum tensor
in terms of pressure and energy density. This issue
must be tackled before we can even write the field
equations.

\subsection{The energy-momentum tensor}

The problem is easily explained as follows: in standard
cosmology one has $T^{00} = \rho_0$ which after
lowering an index using $g_{00} = c^2$, becomes $T^0_{\
0} = \epsilon = \rho_0 c^2$. Here $\rho_0$ is the mass
density in the rest frame of the fluid and $\epsilon$
is the related energy density. This shows that one can
define the energy-momentum tensor in terms of the
``usual'' quantities (density of energy, of mass,
momenta, pressure, etc.) using either contravariant
or mixed indices.

This remains true in VSL theories, even if $\rho$ and
$\epsilon$ do not have the same time dependence
anymore. In this model this arbitrariness is completely
lost. To get through this ambiguity, let us use, as a
guidance, the way the metric tensor is defined. The
metric tensor $g_{\mu \nu}$ is defined in terms of its
covariant components because multiplication by
differentials $\ud x^{\mu} \otimes \ud x^{\nu}$
(product of basis 1-forms) must give the line element
which is a scalar. The other versions, if needed, are
to be derived via the usual indices raising procedure.

By the same token, the components of the
energy-momentum tensor are to be defined in their
contravariant form. This is easy to see in the simple
collisionless case where the energy-momentum tensor is
defined starting from four-velocities in the comoving
frame: $T^{\mu \nu} = \rho_0 \beta^{\mu} \beta^{\nu}$.
Given that four-velocities are four-vectors (defined
with contravariant indices) the energy-momentum tensor
is opportunely defined with contravariant indices as
well. It is only by lowering an index that we can
express $T^{\mu \nu}$ in terms of the energy density of
the fluid. In our model this gives:
\beq
T^0_{\ 0} = \rho \gz^2 =
\frac{\epsilon}{c^2} \gz^2 = 4N^2 \epsilon.
\label{eq:T00}
\eeq
The reason why we have to solve the field equations in
their mixed-indices form is that the energy density
$\epsilon$ must appear in the equations in order to
use the equation of state $w \equiv p/\epsilon$.

Straightforwardly extending what we have just seen to
the case where a pressure is present and the
energy-momentum tensor is defined by 
\beq
T^{\mu \nu} = \left( \rho + \frac{p}{c^2} \right) 
\beta^{\mu} \beta^{\nu} - p g^{\mu \nu},
\eeq
we see that the energy-momentum tensor in the
mixed-indices representation becomes
\beq
T^\mu_{\ \nu} = 
\begin{pmatrix}
4N^2 \epsilon & 0 & 0 & 0\\
0 & -p & 0 & 0\\
0 & 0 & -p & 0\\
0 & 0 & 0 & -p 
\end{pmatrix} = 
\begin{pmatrix}
4N^2 \epsilon & 0 & 0 & 0\\
0 & -w \epsilon & 0 & 0\\
0 & 0 & -w \epsilon & 0\\
0 & 0 & 0 & -w \epsilon
\end{pmatrix}
\label{eq:enmomtens}
\eeq

We stress the great importance of having both $N$ and $w$
appearing in the mixed representation of the energy-momentum
tensor. Without this feature, this model would have had
exactly the same problem that affects standard cosmology and
that has been already discussed: a change in the equation of
state would imply a change in the average curvature of
space-time not deriving from a change in the total
energy-momentum content of the universe, in a flagrant
contrast with the very foundations of General Relativity.

\subsection{Field equations and Bianchi identities}

If one writes the Einstein-Hilbert action for General
Relativity in terms of $x^0$ and variates it, one
finds  that the Einstein field equations are not
affected by the introduction of a varying speed of
light. As a consequence, no additional terms appear in the
field equations which therefore have the usual form.
Using the energy-momentum tensor given by equation
(\ref{eq:enmomtens}), the quantities given in Section
\ref{sec:tensors} and introducing the shorthand notation
\beq
\chi = \chi(t) \equiv \gz
\eeq
we find that the general field
equations for this model are:
\begin{align}
\frac{3}{a^2} \cdot 
\left[ k+ \frac{\dota^2}{\chi^2} \right] &=\,
\frac{8\pi G}{c^4}\: 4N^2 \epsilon 
\label{eq:f1}\\[5mm]
\frac{1}{a^2} \cdot \bigg[k+ \frac{1}{\dota} \, 
\udd \! \left[ \frac{a \dota^2}{\chi^2} \right] \bigg] 
&=\, - \frac{8\pi G}{c^4}\,p =\, 
- \frac{8\pi G}{c^4}\, w\, \epsilon
\label{eq:f2}
\end{align}
where $\epsilon$ is the energy density, $p$ is the
pressure and $w$ is the ratio $p/\epsilon$ representing
the equation of state as usual. 

Imposing the once-contracted Bianchi identities 
$G^{\mu}_{~\nu\, ;\, \mu} = 0$, we see that the usual
conservation law for the energy-momentum tensor is no 
more satisfied: 
\beq
T^{\mu}_{~\nu\, ;\, \mu} \neq 0.
\eeq
The Bianchi identities give instead:
\beq
(k_E\,T^{\mu}_{~\nu})_{;\,\mu} = 0,
\label{eq:Bid}
\eeq
the ``spatial'' equations being trivial as usual,
while the $\nu = 0$ equation for a constant $N \neq0$
is calculated and conveniently rearranged into:
\beq
4N^2\,k_E \left( \frac{\dot{\epsilon}}{\epsilon} +
3(1+\frac{w}{4N^2})\,\frac{\dot{a}}{a} - 
4\, \frac{\dot{c}}{c} + \frac{\dot{G}}{G} 
\right) \epsilon = 0.
\label{eq:conserv}
\eeq

\noindent
Apart from the trivial solution $\epsilon(t) = 0$, the
previous equation has the general solution:
\beq
\epsilon(t) =  \epsilon_0 \cdot
\frac{G_0}{c^4_0} 
\cdot \frac{c^4(t)}{G(t)} \cdot
\left( \frac{a_0}{a(t)} \right)^{3 \left( 1+\frac{w}{4N^2} \right)}
\label{eq:cons}
\eeq
where the index \emph{zero} refers to an arbitrary
time as, for instance, the present time.%
\footnote{As is commonly done in standard cosmology, we
have assumed that $w$ does not vary appreciably over
times where the solution given is used, i.e., we are
studying the dust or the radiation-dominated eras.}

\noindent
If $N=0$, equation (\ref{eq:Bid}) becomes
\beq
3 w \, k_{\!E}\, \epsilon \, \frac{\dot{a}}{a} = 0
\eeq
which is an identity given that $a(t) \propto t^N$.

\noindent
Using equation (\ref{eq:cons}) the field equations become 
\begin{align}
\frac{3}{a^2} \cdot 
\left[ k+ \frac{\dota^2}{\chi^2} \right] &= 
4N^2\, \ek \, \epsilon_0\, 
\left( \frac{a_0}{a} \right)^{3 \left(1+\frac{w}{4N^2} \right)} 
\label{eq:field1}\\[5mm]
\frac{1}{a^2} \cdot \bigg[k+ \frac{1}{\dota} \, 
\udd \! \left[ \frac{a \dota^2}{\chi^2} \right] \bigg]
&=\, -\, w\, \ek\, \epsilon_0\, 
\left( \frac{a_0}{a} \right)^{3 \left(1+\frac{w}{4N^2} \right)}
\label{eq:field2}
\end{align} 
where $\ek$ is the Einstein constant  computed at the
initial conditions imposed upon solving equation 
(\ref{eq:conserv}), its value being:
\beq
\ek = \frac{8 \pi\,G_0}{c_0^4}.
\label{eq:KE}
\eeq

We remark that the evolution of $c(t)$ and $G(t)$ does not
affect the right hand sides of the field equations. Actually
$G(t)$ is completely disappeared from the field equations
(only $G_0$, inside $\ek$, is still present) so that a
possible evolution of $G(t)$ with time has no effect on the
dynamics\footnote{We are assuming that we can introduce a
variable $G(t)$ directly in the field equation. If the
variable $G(t)$ is introduced in the action then the field
equations will be different form those given above. The
results obtained in the paper  remain valid in any case if
one reverts to a constant $G$.}.

We wish now develop further the analysis of the
space-time flat solutions. We will show that indeed
this model admits contracting solutions that can match
observations without requiring the presence of any kind
of exotic matter/energy, i.e., assuming that only
radiation or pressureless matter are present.

\subsection{Solutions from space-time flatness}

Under the requirement of space-time flatness, the most
direct way to obtain information about the solutions to
the field equations is to take advantage of the fact
that the trace of the energy-momentum tensor vanishes:
$T=G=-R=0$. This requirement immediately yields $4N^2 =
3w$, which, solved for N, gives 
\beq
N= \pm \frac{\sqrt{3w}}{2}.
\label{eq:w}
\eeq

Needless to stress how important this relation is, and
what are the implications of the ``$\pm$''. It
directly gives the exponent for the power-law
behaviour of the scale factor in terms of the state
parameter $w$.

Making use of this relation one obtains the following
equations of motion for the scale factor and evolution
for the speed of light:
\begin{align}
& a(t) \,= \,a_0\, 
\left( \frac{t}{t_0} \right)^{\pm \sqrt{3w}/2} 
\label{eq:aditw}\\
& c(t) \,= \,c_0\, 
\left( \frac{t}{t_0} \right)^{\pm \sqrt{3w} -1}.
\label{eq:cditw}
\end{align}
These are to be compared to the solution one 
obtains for the standard FLRW universe:
\begin{align}
& a(t) \,= \,a_0\, 
\left( \frac{t}{t_0} \right)^{2/3(1+w)}\\
& c(t) \,= \,c_0\,.
\end{align}

Let us now analyse the properties of the solutions we have
just found. The first feature one notes is that we recover
the same solution for the expanding radiation-dominated era,
$a(t) \propto t^{1/2}$, but also obtain the solution $a(t)
\propto t^{-1/2}$, which is the contracting solution we were
looking for. As for the matter era, we find $a(t) =
\textsc{constant}$. We remark however that the constant
solution is exactly constant if and only if nothing but
pressureless dust contributes as source into the
energy-momentum tensor. Even the smallest contribution from
any form of energy (e.g.~radiation) different from the pure
dust is sufficient to give a varying scale factor.
Therefore, this static solution is just an idealized and
asymptotic one. 

The crucial point to note is that the general solution given
in equation (\ref{eq:w}) present us with two solutions, one
being the reciprocal of the other. The most evident
implication is that a universe containing any given mixture
of dust and radiation (with one dominating over the other)
can be either in a contracting or in a expanding motion, and
which of this two motions is actually taking place cannot be
told by estimating the content of the universe, because the
same $w$ is compatible with both solutions. So, for any
given equation of state, one always has two possible
solutions for the state of motion of the universe, a
contracting one and an expanding one.

On observational grounds, one could think that the
motion effectively experienced by our universe is
that of an expansion, but one would then be forgetting
Table \ref{tab:evolution}, which tells us that both an
expansion and a contraction are perceived as an
expansion by a constant-$c$ observer. Indeed, the
apparent first derivative of the scale factor,
$\dot{A}$, is positive for \emph{any} $N$. But if both
``thermodynamical'' and ``kinematical'' observations
cannot tell an expanding from a contracting universe,
does it make any sense at all to distinguish between
the two solutions?

The answer is that there actually is a kinematical way
to tell one solution from the other, but it is not
related to the first derivative. Rather, the imprints
of a contraction are to be read from the apparent
second derivative. Looking again at Table
\ref{tab:evolution}, we see that if $N<0$ or $N>1$ then
the apparent second derivative turns out to be
positive, while for $0<N<1$ it is negative. 
Observations tell us that $\ddot{A} > 0$, so that our
universe is now a mixture of dust and radiation-like
matter and is undergoing an accelerated contraction
powered by energy-pressure. \emph{That is how the
puzzle of the acceleration of the universe becomes the
only instrument we have to tell in which phase,
expanding or contracting, the universe presently
happens to be.}

\subsection{An eternally bouncing universe?} 

It is interesting to speculate on what the global history of
such a universe could have been and what its future may be.
A possible scenario is that of a universe undergoing an
eternal cycle of successive expansions and contractions.
This is similar to what happens in the standard FLRW
universe with a positive spatial curvature, but with the two
main differences that here the universe is both space-time
and spatially flat, and that the evolution law for the scale
factor is not a cycloid but rather a power law with a
varying exponent.

To give a brief qualitative description of such possibility,
let us start from a hot and dense universe where radiation
dominates. Once the radiation era starts, the universe
expands according to the well known power law $a(t) \propto
t^{1/2}$, with the exponent decreasing as the expansion
cools the universe until pressureless matter dominates over
radiation. In this phase the expansion becomes very slow
(the exponent tends to zero). 

Since pressure is ever decreasing, general relativity
approaches a Newtonian behaviour, where mass density
generates what looks like as an attractive force. For some
reason (e.g.~because of substantial deviations from a pure
homogeneous and isotropic scenario), the expansion might
eventually stop and an almost-classical contracting motion
could begin. Such motion is described by the same power law
but with an exponent of opposite sign. According to this
almost-Newtonian behaviour, the contraction accelerates more
and more (exponent in the power law increasing more and more
as the universe contracts), until the pressure regains his
role in the game. 

Eventually, the universe enters a new radiation-dominated
era and, according to our solution, the exponent $N$
stabilizes to a value equal to $-1/2$. The now constant but
steady contraction will finally cause the matter density and
the pressure to reach such high values that quantum effects
become important and classical general relativity loses its
fundamental character. Possibly, during this quantum era,
the universe may experience a sort of Big Crunch followed by
a new Big Bang which makes the cycle to start again.

\subsection{Evolution of $a(t)$ and $c(t)$ and some
related speculations}
\label{sec:aacc}

Given that $g_{tt} = \chi^2 = 4N^2 c^2(t)$, that
equation \eqref{eq:w} implies 
\beq
w \in [0;1/3] \, \longrightarrow \,
N \in [-1/2;1/2]
\eeq
and that 
\[ 
c(t) = c_0 \left( \frac{t}{t_0} \right)^{2N-1},
\] 
we find that both $g_{tt}$ and $c(t)$ are non-increasing
functions of $t$ for any  $N \in [-1/2;1/2]$. In Table
\ref{tab:ac} a summary of the asymptotic evolutions of
$c(t)$ and $a(t)$ in both the contracting and the expanding
phases is given.

\begin{table}[!h]
\begin{center}
\begin{tabular}{|c||c|c|c|c|}
\hline
& \multicolumn{2}{|c|}{$a(t)$} & 
\multicolumn{2}{|c|}{$c(t)$}\\
\hline
& Expansion   & Contraction   & Expansion & Contraction\\
\hline
\hline
$w=0$ & $t^0$ & $t^0$ & $t^{-1}$ & $t^{-1}$\\
\hline
$w=1/3$  & $t^{1/2}$ & $t^{-1/2}$ & $t^0$ & $t^{-2}$\\
\hline
\end{tabular}
\end{center}
\caption{Evolution of $a(t)$ and $c(t)$ for $w=0,1/3$.}
\label{tab:ac}
\end{table}

An interesting feature is that, for the radiation-dominated
era in the expansion phase, we recover the standard FLRW
cosmology: $a(t)$ evolves as $t^{1/2}$, $4N^2\!=\!1$ and
$c(t)$ is constant. The FLRW model matches then with our
model during a small but important part of the history of
the universe. Most importantly, this guarantees that the
description of the post-inflationary early phases of the
universe in our model is precisely the same of the classical
FLRW description. For instance, results concerning the BBN are
not affected by a varying-$c$ as $c(t)$ does not vary at
those high energies in the present model. Note that this is
the opposite of what happens in other popular VSL models,
where the speed of light often varies only at very high
energies (during a phase transition).

Apart from this brief period where $c(t)$ is
asymptotically constant, the \ls is a decreasing
function of time. $c(t)$ varies according to a power
law with the exponent ranging from the idealized value
of 0, during the radiation-dominated expansion phase,
to a maximum value of -2, during the
radiation-dominated contraction phase. This guarantees
that any given part of the universe was causally
connected to any other different part at some moment in
the past. No matter how far two points in space are
nowadays, one can always find an early enough time when
these two points were causally connected. \emph{In this
model there is no horizon problem.}

Another curiosity we wish to address is that for $N=\! \pm
1/2$ -- that is, during the radiation-dominated eras -- we
have $\chi(t)=c(t)$ so that one recovers the ``usual'' line
element $\ud s^2 = c(t)^2 \ud t^2 - a(t)^2 \vec{\ud r}^2$
(usual in the sense that $c(t)$ appears instead of
$\chi(t)$). This suggests the possible existence of an
interconnection between the fact that photons (massless
particles) travel along null geodesics, the fact that they
travel precisely at a speed equal to $c(t)$ and the fact
that the pressure of radiation is equal to 1/3 of its
energy density. In fact, if any of the previous things is
changed, then the other two will change as well: all three
are concomitantly true or false. Such result suggests that
some properties of massless particles may have a
cosmological origin. 

The dust-dominated era is not less intriguing: given
that $\chi = 4N^2 c(t)^2 \to 0$ when $N \to 0$, we
have a sort of ``cosmological indication'' that
massive particles have smaller speeds with respect to
the locally measured speed of light. It is tempting to
use na\"ive kinetic theory to transform the line
element. From the usual formula: 
\beq
p = \frac{1}{3} \frac{\epsilon}{c^2} (v_{\textsc{RMS}})^2
\eeq
we obtain
\beq
w = \frac{1}{3c^2} (v_{\textsc{RMS}})^2
\eeq
that, once substituted into $g_{tt}=4N^2c^2(t)$, making
use of $N^2=3w/4$, gives
\beq
g_{tt}= (v_{\textsc{rms}})^2
\eeq
so that the line element becomes
\beq
{\ud s}^2 = v^2_{\textsc{rms}}(t) {\ud t}^2 - 
a(t)^2 {\vec{\ud r}}^2
\eeq
which seems to suggest that particles of any mass,
depending on the idealized content we attribute to the
universe, can define worldlines of null length.

Of course we are abusing here of the concept of
$v_{\textsc{rms}}$ by assuming that any particle in the
swarm has the same velocity. If we however were to take all
the particles to have the same velocity (a very highly
idealized application of the cosmological principle), we
could drop the \textsc{rms} index and what has been
conjectured above would become a little bit more precise.

This discussion is of course highly speculative and would
only make sense in a hypothetical universe where one and
only one sort of fluid were present and where all the
particles of that kind were to travel at the same speed.
Consequently, the remarks made above are in no way not
related to our universe. However, some interesting effects
could take place even when a more realistic case is
considered and this is why I have mentioned these
reflections here.

\subsection{Initial conditions}

We now solve the field equations directly to look for
constraints on initial conditions.
We recast equation (\ref{eq:const}) in the form
\beq
\frac{\dot{a}}{\chi} \, = 
\, \frac{a^2_0}{2t_0c_0} \, \frac{1}{a}
\eeq
and use it in equation (\ref{eq:field1}) with $k=0$ to
obtain
\beq
\frac{3}{a^4} \, \left( \frac{a^2_0}{2t_0c_0}
\right)^2  = \,4N^2 \, \ek \, \epsilon_0\,  \left(
\frac{a_0}{a} \right)^{3 \left(1+\frac{w}{4N^2}
\right)}. 
\label{eq:fe1} 
\eeq

We see that $a(t)$ drops out form this equation if and only
if $3w=4N^2$, which is precisely the solution to
equation (\ref{eq:w}). In this case simplifications
yield
\[
3=(4Nt_0c_0)^2 \ek \epsilon_0
\]
that can be recasted, for instance, into
\beq
t_0 \, = \frac{1}{4N c_0} \, 
\sqrt{\frac{3}{\ek \epsilon_0}}
\label{eq:t0}
\eeq

Using the accepted numerical values referred to the
present time for the quantities involved (assuming
that such estimations are not excessively model
dependent, which is far from being granted) one
obtains  
\beq
t_0 \simeq \frac{10^{26 \div 27}}{N} \ \textrm{sec}
\label{eq:t0n}
\eeq
the interpretation of which cannot of course be
straightforward for a variety of reasons. 

For instance, with $g_{tt}$ depending on time, the
cosmological clock is seen to tick slower or faster
depending on the epoch the observer
happens to live in\footnote{For instance, if an
observer at rest in a radiation-dominated area of the
universe were able to observe the clock of another
observer at rest in the same universe, but in a far
enough portion of it which happens to be
dust-dominated, then he would observe that clock
ticking slower with respect to his own clock.}. This is
reflected by the presence of $N$ in the denominator of
the previous equation. In fact, for $N \to 0$ (matter
era), $t_0 \to \infty$ in agreement with the fact that
in this model the cosmological clock is seen to tick
slower during the matter era ($g_{tt} \to 0$ when $N
\to 0$).

Moreover, $g_{tt}$ contains the \ls $c(t)$, which has been
showed to be a non-increasing function of time. This implies
that, apart form the effect discussed above, the
cosmological clock ``ticked faster in the past'' so that, in
a sense, $t_0$ is expected to be much smaller than the value
given in equation (\ref{eq:t0n}) From this qualitative
discussion it emerges that the value given above cannot be
directly compared to other estimations of the age of the
universe based upon a constant \ls (and a single-cycle
universe).

Note that in the case $3w \neq 4N^2$, equation
(\ref{eq:fe1}) reduces to 
\beq
\frac{a_0}{a} = 
\left[ \frac{(4N t_0 c_0)^2 \ek \epsilon_0}{3}
\right]^{\frac{4N^2}{3w-4N^2}}
\eeq
which implies that $a(t) = \textsc{constant}$ and
therefore $a(t) = a_0$ so that again equation
(\ref{eq:t0}) holds. However, equation (\ref{eq:const}),
that we have used to simplify (\ref{eq:fe1}), follows
from the hypothesis of space-time flatness ($R=0$), that is
from $3w = 4N^2$, so that one could hardly give any
meaning to the $3w \neq 4N^2$ case.

We turn our attention to the second field equation
(\ref{eq:field2}). Using the second of the 
expressions given for $G^r_{~r}$, we get
\beq
\frac{1}{a^2} \left[ k+ \frac{1}{a \dota}\, \udd 
\left[ \left( \frac{a \dota}{\chi} \right)^2 \right] -
\frac{1}{a^2}\, \left( \frac{a \dota}{\chi} \right)^2 
\right] = \, - w\, \ek \, \epsilon_0 \, 
\left( \frac{a_0}{a} \right)^{3 \left(1+\frac{w}{4N^2}
\right)}
\eeq 
where, by equation (\ref{eq:const}) with $k=0$, the
first and the second terms on the left hand side vanish
while the last term is constant. Under these
conditions one can recast the second field equation
into
\beq
\frac{-1}{a^4}\, 
\left( \frac{a_0^2}{2 t_0 c_0} \right)^2\, =
\, -w\, \ek \, \epsilon_0 \, 
\left( \frac{a_0}{a} \right)^{3 \left(1+\frac{w}{4N^2}
 \right)}
\eeq 
which both in the case $3w=4N^2$ and $3w \neq 4N^2$
gives the same result -- equation (\ref{eq:t0}) --
obtained from the  first field equation. 

The fact that the field equations do not give
additional information about the dynamics was to be
expected as we have applied two constraints --  space-time
flatness and the Bianchi identities -- on a system with
two degrees of freedom: the two scale factors $a(t)$
and $\chi(t)$. In such situations both the field
equations become ``initial conditions'' equations and,
as is to be expected, both give the same result.


\section{Consistency checks and predictions}
\label{sec:conschecks}

The new VSL model presented in this paper is capable of
solving the puzzle of the acceleration of the universe but
some of its aspects, such as the presence of dust and
radiation only or that the acceleration is only an apparent
effect must surely result somewhat unfamiliar and one could
wonder if the model is compatible with the relevant
observative evidences. While a complete data analysis
related to the most important observative constraints is
evidently beyond the scope of the paper, some consistency
checks are perhaps at order, at least to be sure that no
contradictions, either intrinsic or related to observations,
could arise. We do not digress on the SN data as it is
evident that the presence of an apparent acceleration makes
our model compatible with that data set. On the contrary,
there are two issues which are to be addressed with care:
the critical density for the model, which is an internal
consistency check, and, as anticipated at the end of section
\ref{sec:appacc}, the relation with the variation of
$\alpha_\textsc{em}$, which can be considered the only
observative evidence which is somewhat related to a
varying-$c$.

\subsection{Critical density}

We have already emphasized that only dust and radiation
are needed for our model to exhibit an (apparent)
accelerated expansion. No dark matter\footnote{Dark
matter is not excluded as a further  contribution to
the energy-momentum tensor for the model.}, let alone
dark energy, are needed to explain the acceleration of
the universal expansion, which was the main issue we
wished to address. However, in our analysis we have
also assumed that the universe is spatially flat and
this implies that the energy-mass density must be
critical according to equation (\ref{eq:f1}) (with $k$
set equal to zero) that we re-express for convenience as
\beq
\rho_{\textsc{cr}} = \frac{H^2}{24\pi G w^2}\,.
\label{eq:crd}
\eeq

This critical density must be achieved without invoking
exotic forms of energy as these have not been introduced
into the model. At first this may seem to be a fatal
shortcoming for the model as the average baryonic plus
relativistic matter-energy density of the universe is
observed to be around the 4\% of the standard-cosmology
critical density $\rho^{\textsc{flrw}}_{\textsc{cr}} =
3H^2/8 \pi G$ implied by the standard Friedmann equation.
Things worsen if one considers that, for a
dust-dominated universe,  $\rho_{\textsc{cr}}$
given by equation (\ref{eq:crd}) is greater  than
$\rho^{\textsc{flrw}}_{\textsc{cr}}$. 
Indeed one finds that
\beq
\rho_{\textsc{cr}} = \frac{1}{(3w)^2}\,
\frac{H^2}{H^2_\textsc{flrw}} \, 
\rho^{\textsc{flrw}}_{\textsc{cr}}.
\eeq
From
\beq
a(t)_\textsc{this model} \,= \,a_0\, 
\left( \frac{t}{t_0} \right)^{\pm \sqrt{3w}/2}\, ; \qquad
a(t)_\textsc{flrw} \,= \,a_0\, 
\left( \frac{t}{t_0} \right)^{2/3(1+w)}
\eeq
follows
\beq
\frac{H^2}{H^2_\textsc{flrw}} = 
\frac{3w(w+1)^2}{16}
\eeq
and therefore
\beq
\rho_{\textsc{cr}} \,=\, 
\frac{3(w+1)^2}{16w} \,\rho^{\textsc{flrw}}_{\textsc{cr}}.
\eeq

The last relation shows that $\rho_{\textsc{cr}} =
\rho^{\textsc{flrw}}_{\textsc{cr}}$ for a universe filled
with radiation\footnote{This further confirms that, as we
had found previously, the FLRW model and this model
are indistinguishable during the expanding 
radiation-dominated era.} but $\rho_{\textsc{cr}} >
\rho^{\textsc{flrw}}_{\textsc{cr}}$ when dust dominates.
After a closer inspection however, it becomes clear that
things are a little more complicated. In fact, the critical
density is calculated from the estimations made on the
Hubble parameter $H \equiv \dota/a$ which, however, has an
apparent counterpart $H_{\textsc{app}} \equiv \dot{A}/A$
whose interpretation should be obvious by now: it is
$H_{\textsc{app}}$ and not $H$ that is inferred from
measurements and is therefore incorrectly used to determine
what indeed is an \emph{apparent critical density}. By means
of $H$, we must now calculate the real critical density
which is what is to be compared with other
(redshift-independent) estimations of the matter density to
check whether this model is consistent with observations or
if it necessitates a generalization to include other forms
of matter-energy in order to cope with observative data. 

A convenient way to follow is to calculate the ratio between
the critical density as it appears from observations and the
real critical density, that is, the matter density which is
actually needed to have a spatially flat universe. The
computation is straightforward and yields, for the present
time\footnote{For arbitrary times the solution is
$3w(t_0/t)^{2(\sqrt{3w}-1)}$.}, 
\beq
\frac{\rho_{\textsc{cr}}}{\rho^{\textsc{app}}_{\textsc{cr}}}
= 3w 
\label{eq:wpred}
\eeq
whose order of magnitude may range from $10^{-1}$ to
$10^{-4}$ depending on several issues as, for instance,
neutrino contribution to the present-day energy density and
pressure. In other words, the true critical density is a
factor $10^{-1}$ to $10^{-4}$ smaller than the estimated
value. This broad range is in a qualitative agreement with
the above-mentioned 4\% discrepancy. 

Summing up, we have found that a determination of the
critical density made by estimating the Hubble parameter
leads to a large overestimation during the matter era. Much
less mass-energy is needed to have a spatially flat universe
with respect to that suggested by the measure of the Hubble
parameter. This finding makes the model internally
consistent and provides an elegant solution to the
problem of the missing mass/energy with no need to resort
to hypothetical forms of matter/energy. \emph{We have
constructed a spatially-flat, ``normally''-populated,
apparently-accelerating universe.}

\subsection{Present-time density contribution from 
relativistic matter}

Equation \eqref{eq:wpred} predicts the existence of a
contribution from relativistic matter to the present-time
matter density and allows to obtain a rough prediction of
its value. This contribution is  necessary for the model to
be self-consistent. We assume that there is no need to
introduce any dark matter into the model and denote with
$\Omega_\textsc{n}$ the \emph{observed} fraction of the
``normal''  matter density $\rho$ with respect to the
\emph{actual} critical density $\rho_{\textsc{cr}}$. The 
currently accepted value for $\Omega_\textsc{n}$ is 
$\Omega_\textsc{n} = 0.04$. If the universe is spatially 
flat the observed matter density must be equal to the 
apparent critical density, so that
\beq
\rho_{\textsc{obs}} = \rho^{\textsc{app}}_{\textsc{cr}}
\eeq
must hold. From equation \eqref{eq:wpred} it follows then
that it must be
\beq
3w = \Omega_\textsc{n}
\label{eq:omegan}
\eeq
whence 
\beq 
w_\textsc{today} \,\simeq\, 0.013.
\label{eq:wtd}
\eeq

Given that for relativistic matter
$p_\textsc{rel}=\epsilon_\textsc{rel}/3$, we find
\beq
w \,=\,\frac{p}{\epsilon} \,=\, 
\frac{\epsilon_\textsc{rel}}{3 \epsilon_\textsc{tot}}
\eeq
which, combined with equation \eqref{eq:omegan}, gives
\beq
\frac{\epsilon_\textsc{rel}}{\epsilon_\textsc{tot}} 
\,=\, \Omega_\textsc{n}
\eeq
The last equation tells us that \emph{the present-time
contribution of relativistic matter to the total matter
density is predicted by the model to be of order 4\%}. Such
contribution must originate from ``normal'' (i.e.\ well
known and observed) relativistic matter and the optimal
candidate for this role is the neutrino. Given the
uncertainty about neutrino masses this 4\% contribution is
well within the observative bounds. Actually, the previous
result may be of help in determining neutrino masses and
number densities from cosmological measurements but a more
extensive analysis is beyond the scope of this paper.

\subsection{Connection with the $\alpha_{\textsc{em}}$
results}

As anticipated at the end of section \ref{sec:appacc}, the
existence of the apparent effects discussed throughout the
paper relies on the assumption that the
$\alpha_{\textsc{em}}$ results do not have a large impact on
the evolution of emission frequencies. In this section we
discuss this issue and show that our results are correct 
whatever the evolution of $\alpha_{\textsc{em}}$. We also
discuss whether the variation of $\alpha_{\textsc{em}}$ can
be explained in terms of the variation of $c$ only. As we
are about to see, this problem is in general ill-posed but
if we specialize to the units/frame we have adopted, we find
that the variation of $\alpha_{\textsc{em}}$ induced by that
of $c$ is much larger that the observed one. This finding
forces us to conclude that in our case $e$ and/or $h$ must
also vary. Note that this fact has no impact on the results
that we have obtained so far as these are not affected by
the evolution of $h$ and/or $e$.

We first point out that, contrary to what one may be led to
think at a first glance, the variations of $c(t)$ and of
$\alpha_{\textsc{em}}$ do not necessarily imply each other.
Indeed one may think that, since $\alpha_{\textsc{em}} =
e^2/hc$, if $c$ varies a variation of $\alpha_{\textsc{em}}$
is to be expected as well. This first impression is wrong.
As recalled in the introduction, $\alpha_{\textsc{em}}$ is a
dimensionless quantity so that its variation (or
non-variation) is independent from the units chosen. On the
contrary, whatever the behaviour of $\alpha_{\textsc{em}}$,
one may choose the units/frame so that $c$ (and/or
$e$ and/or $h$) vary or remain constant\footnote{Of course,
if $\alpha_{\textsc{em}}$ is constant, a variation of $c$
must be compensated by a variation of $e$ and/or $h$.}.

Given that $\alpha_{\textsc{em}} \propto e^2/hc$, we have
\beq
\frac{\dot{\alpha}_{\textsc{em}}}{\alpha_{\textsc{em}}}
\,=\;
2\frac{\dot{e}}{e} - \frac{\dot{c}}{c} - 
\frac{\dot{h}}{h}
\eeq
which, if the dependence on time (or redshift) is linear,
becomes
\beq
\frac{\Delta \alpha_{\textsc{em}}}{\alpha_{\textsc{em}}}
\,=\;
2\frac{\Delta e}{e} - \frac{\Delta c}{c} - 
\frac{\Delta h}{h}
\label{eq:aech}
\eeq
From this relation one sees that $\alpha_{\textsc{em}}$ may
remain constant and nonetheless $c$ may be time dependent,
if its variation is compensated by that of $e$ and/or of
$h$. Vice versa, $c$ can be set to a constant by an
appropriate choice of units/frame while
$\alpha_{\textsc{em}}$ could vary as a consequence of the
variation of $e$ and/or of $h$. This remains true in
general, even if the dependence is not linear. It is because
of these arguments that we affirmed that the question
whether the variation of $\alpha_{\textsc{em}}$ can be
explained only in terms of the variation of $c$ is
ill-posed: it all boils down to the choice of
units/frame\footnote{Some may consider this argument 
self-evident but there are still physicists who ask  whether
the observed evolution of $\alpha_{\textsc{em}}$ is
compatible with this or that model. They do not realize that
the point is meaningless.}. Specifying some units/frame
gives a meaning to the issue and we will verify what happens
in the units/frame adopted here  at the end of this section.

A somewhat related issue is the influence of a
varying-$\alpha_{\textsc{em}}$ on the results that we have
obtained so far: can a varying-$\alpha_{\textsc{em}}$ affect
our findings in some way? As explained in section
\ref{sec:postulates}, we have adopted the frame in which the
emission/absorption frequencies, as seen by an observer at
rest with the cosmological fluid, are the same at any cosmic
time. It is not difficult to see that this remains strictly
true even if $\alpha_{\textsc{em}}$ is evolving with
cosmological time. This is because the emission/absorption
frequencies are dimensional quantities and have the
dimension of the inverse of time. It follows that, even if
$\alpha_{\textsc{em}}$ is a factor appearing in the
theoretical expression for the emission/absorption
frequency, it must appear multiplied by an appropriate 
dimensional factor in such a way to recover the correct
dimension. Such a combination will therefore be dimensional
and as such will depend on the units/frame adopted: even if
$\alpha_{\textsc{em}}$ is present, one will always be able
to fix the overall expression for the emission/absorption
frequency equal to a constant. This is precisely what we
have done upon choosing the frame adopted in the paper. The
crucial consequence of these arguments is that \emph{any
evolution or non-evolution of $\alpha_{\textsc{em}}$ is
compatible with our model}. Alpha results alone can in no
way confirm or falsify it.

We next examine what is the variation of
$\alpha_{\textsc{em}}$ induced by that of $c(t)$ obtained
in the paper. We assume for the moment that $h$ and $e$ are
constant and will check this assumption against the
$\alpha_{\textsc{em}}$ results. 
We denote the said quantity with $\left( \Delta
\alpha_{\textsc{em}}/\alpha_{\textsc{em}} \right)_c$\ :
\beq
\left( 
\frac{\Delta \alpha_{\textsc{em}}}{\alpha_{\textsc{em}}} 
\right)_c
\equiv\;
\frac{\Delta \alpha_{\textsc{em}}}{\alpha_{\textsc{em}}}
\ \ \textsc{induced by}\  c(t).
\eeq

Assuming that the variation is almost linear would be too
rough an approximation so that we do no use equation
\eqref{eq:aech} and calculate the evolution directly.
Observative results are given in terms of $\Delta
\alpha_{\textsc{em}}/\alpha_{\textsc{em}}$ which is defined
as 
\beq
\frac{\Delta \alpha_{\textsc{em}}}{\alpha_{\textsc{em}}}
\equiv
\frac{\alpha_{\textsc{em}}(z) - \alpha_0}{\alpha_0}
\label{eq:dadef}
\eeq
where $\alpha_0 \simeq 1/137$ is the present-time value of the 
fine-structure constant while $\alpha_{\textsc{em}}(z)$ is
the value it assumes at redshift $z$. 
With $e$ and $h$ constant $\alpha_{\textsc{em}} \propto 1/c$
so that we obtain
\beq
\left(
\frac{\Delta \alpha_{\textsc{em}}}{\alpha_{\textsc{em}}} 
\right)_c \,=\, \frac{c_0}{c(t)}-1.
\eeq
Using equation \eqref{eq:cditw} we find that\footnote{In
equation \eqref{eq:cditw} we are taking the minus-sign
solution because in the present time the universe is 
undergoing a contracting motion.}
\beq
\left(
\frac{\Delta \alpha_{\textsc{em}}}{\alpha_{\textsc{em}}} 
\right)_c \,=\;
\left( \frac{t}{t_0} \right)^{1+\sqrt{3w}} -1
\label{eq:acprel}
\eeq
In equation \eqref{eq:dadef} $z$ is the \emph{observed} $z$
and therefore it is an apparent quantity which we will
appropriately denote with $z_\textsc{app}$ and which is
obviously to be defined\footnote{Recall that what instruments
measure is  a frequency shift, see discussion in section 
\ref{subsec:anintarg}.} as
\beq
1+z_\textsc{app} \equiv \frac{A_0}{A(t)}.
\eeq
We then need a link between $z_\textsc{app}$ and $t$. 
From equation \eqref{eq:ascafa}, recalling that $N=1-S$, we 
find that $A(t) \propto t^{1-N}$ and therefore
\beq
1+z_\textsc{app} = \left( \frac{t}{t_0}
\right)^{-1-\frac{\sqrt{3w}}{2}}
\eeq
where we have used the contracting solution for equation 
\eqref{eq:w}.
Inverting the previous relation and using equation
\eqref{eq:acprel} we conclude that
\beq
\left(
\frac{\Delta \alpha_{\textsc{em}}}{\alpha_{\textsc{em}}} 
\right)_c \,=\, 
\left( 1+z_\textsc{app} 
\right)^{-2\frac{1+\sqrt{3w}}{2+\sqrt{3w}}}-1 
\label{eq:acdef}
\eeq
which can also be expressed as
\beq
\alpha_{\textsc{em}}(z_\textsc{app})
\,=\, \alpha_0
\left( 1+z_\textsc{app} 
\right)^{-2\frac{1+\sqrt{3w}}{2+\sqrt{3w}}}.
\eeq
This is the variation of $\alpha_{\textsc{em}}$ 
induced by the variation of $c(t)$ obtained in the paper.

Note that we have assumed throughout that $w$ is constant
over the time interval where the measuraments have been
performed. This approximation is justified as follows.
Using equation \eqref{eq:wtd}, we find that
\beq
-2\frac{1+\sqrt{3w_\textsc{today}}}{2+\sqrt{3w_\textsc{today}}}
\simeq -1.1.
\eeq
Going backward in time in our contracting universe means
approaching the asymptotic value $w_\textsc{mat}=0$ for which
one gets
\beq
-2\frac{1+\sqrt{3w_\textsc{mat}}}{2+\sqrt{3w_\textsc{mat}}}
= -1.
\eeq
It then follows that the assumption $w$=\textsc{constant} is
well founded. 
Given that in this section we have used the contracting
solution, the results presented are only valid as far as one
looks back no further than the start of the contracting
phase.

If for $z_\textsc{app}$ we use the mean of the redshifts
used in \cite{Mur2}, that is $z_\textsc{app} \simeq 1.75$,
and again equation \eqref{eq:wtd}, we see that the variation
of $c(t)$ alone would imply a much greater variation of
$\alpha_{\textsc{em}}$ with respect to the observation:
\beq
\left(
\frac{\Delta \alpha_{\textsc{em}}}{\alpha_{\textsc{em}}} 
\right)_c \,\simeq\, 
\left( 2.75 \right)^{-1.1}-1
\,\simeq\, -0.67
\eeq
or
\beq
\alpha_{\textsc{em}}(z_\textsc{app}=1.75)
\,\simeq\, 0.33\,\alpha_0\, .
\eeq
Comparing this result with the observed value \cite{Mur2}
\beq
\left(
\frac{\Delta \alpha_{\textsc{em}}}{\alpha_{\textsc{em}}} 
\right)_\textsc{obs} \,=\, (-0.57 \pm 0.10)\cdot 10^{-5}
\eeq
we conclude that, in the units/frame adopted, $h$ and/or $e$
must also depend on time. We stress that this fact has no
impact on the cosmological results we have obtained in the
paper as none of them is affected on the evolution of $h$
and/or $e$.


\section{Summary}
\label{sec:conclu}

In this paper we have developed a new varying speed of
light model and have attempted to find an explanation
for the acceleration of the cosmological expansion
which does not have to resort to the introduction of 
exotic (and so far unobserved) forms of energy or of a
cosmological constant. 

The model is based on the reconsideration of the most
general form of the metric when the speed of light is
allowed to vary. We have found that, if we want to ascribe
the observed redshifts to and only to the dynamics of the
space, it is convenient to work with a particular time
dependent $g_{tt}$. Appealing to general covariance to
perform a coordinate transformation and obtain a metric with
a constant speed of light, while of course possible, has the
drawback of destroying the above-mentioned property. On the
contrary, working with the units/frame employed in the paper
ensures that any given emission/absorption frequency is the
same at any given epoch of the past and today, which clearly
is the prerequisite to directly compare frequency shifts
and to intepret them as purely cosmological in origin.

We have argumented that the space-time curvature of the
universe, here represented by the Ricci scalar, cannot have
changed during its history. Na\"ively, this can be justified
because, by definition, the universe has nothing
``external'' to interact with, and therefore cannot exchange
energy-momentum.  As a particular case, we have assumed that
the Ricci scalar vanishes. Our choice is dictated by the
fact that the classical universe is believed to emerge from
the quantum universe in a radiation-dominated state so that
the trace of the energy-momentum tensor, and therefore the
Ricci scalar, must vanish, whatever the model. This is
equivalent to saying that, in a sense, the classical
universe has come into existence with no net space-time
curvature. It is to be noted that the same requirement
cannot be made in the framework of standard FLRW cosmology
as it would correspond to imposing the equation of state for
radiation at any time, which would clearly contradict the
observations. In our model, instead, this is not a problem
because fixing the Ricci scalar does not fix the equation of
state.

We have shown the existence of the possibility that the
observed acceleration of the cosmic expansion is an apparent
effect induced by the observer not taking the varying speed
of light into account. This effect arises because
instruments measure frequencies while the dynamics of the
universe is represented by the evolution of wavelengths and
the former does not evolve as the reciprocal of the latter
if the speed of light varies with time.

Upon deriving and solving the new field equations for the
spatially-flat case we have found that both $a(t)$ and
$c(t)$ evolve according to a power law. For any admissible
matter content -- in our case for any $w \in [0,1/3]$ where
$w$ is the usual equation of state parameter $w=p/\epsilon$
-- the new field equations have both an expanding and a
contracting solution and the two solutions are one the
reciprocal of the other. More precisely, we have found that
the scale factor evolves according to $a(t) \propto t^N$
where $N\! = \pm \sqrt{3w}/2$.

At a first glance, a contracting solution seems to
contradict the fact that we observe redshifts and not
blueshifts but it is not so because at fixed $w$ the model
predicts that an observer who is not taking the varying
speed of light into account will observe the frequencies
redshifted exactly by the same amount for both the expanding
and the contracting solutions. This implies that the
contracting solution is not ruled out by observations but,
as a consequence, it also means that the fact that redshifts
and not blueshifts are observed is not sufficient to
discriminate between the two possibilities. Fortunately, the
sign of the apparent acceleration is found to be different
for the two mentioned states of motion. Thanks to this fact,
we have been able to conclude that the observed acceleration
implies that the universe is presently undergoing a
contracting motion. That is how one of the most tantalizing
puzzles of modern cosmology, the accelerating universe,
might turn out to be the only way we have to discriminate
between the two possibilities and determine the actual state
of motion for the universe.

Some consistency checks have been performed. We have found
that the new critical density is inside the observative
bounds for dust plus relativistic matter density, and
therefore there is no need to add hypothetic forms of
matter/energy. We have also verified that in the units/frame
adopted, the evolution of $c(t)$ implies a variation of
$\alpha_\textsc{em}$ which is much larger than the observed
one. Contrary to the common belief, we have shown that this
does not represent a problem but just implies that, in the
units/frame adopted, either $e$ or $h$ or both must evolve
with cosmological time together with $c(t)$. As $e$ and $h$
have no role in our computations this fact in no way affects
our results. 

We remark that our findings have been obtained by
considering the presence of dust and radiative matter only.
No exotic forms of dark matter/energy or even a cosmological
constant are to be introduced for the model to reproduce the
observed evolution. This appears to be a very important
feature as, despite the enormous efforts and investments,
none of the mentioned forms of matter have been directly
observed so far, nor has the cosmological constant received
a satisfactorily theoretical explanation. 

Now we have a VSL model which describes a homogenous,
isotropic, spatially-flat, ``normally''-populated and
apparently-accelerating universe. Such a model is clearly
interesting by itself but there can be little doubt that it
can be improved in a variety of ways. The cases of a
spatially-curved universe and that of a double-component
(dust \emph{and} radiation) fluid are its natural extensions
and we will investigate such possibilities in forthcoming
research work.


\section*{Acknowledgement}
I am deeply indebted to Prof.~Silvio Bergia for his
support, for the useful discussions we have had while this work
was in progress and for critically reviewing the paper.

\appendix

\section{Proportionality between the scale factor and
wavelengths}

Doubts have been posed as to whether the scale factor is
proportional to wavelengths in the model proposed as it is
in standard cosmology. Even at a conceptual level these
doubts sound unfounded as the dynamics of the scale factor
is nothing but the stretching of space. This is precisely
what causes the stretching of wavelengths so the two
quantities evolve in the same way just because, in a sense,
they are the same thing. At any rate, in this appendix we
provide a formal way to see this.

Consider the travel of a light ray from a distant object to
an Earth-based instrument and two successive crests denoted
with 1 and 2. The space separation between the two crests is
one wavelength and the difference of emission times is the
inverse of the emission frequency. The relation linking the
two quantities at the time of emission as well as at the
time of reception is $\lambda(t) \nu(t) = c(t)$. Instant by
instant a given light ray will follow a null geodesic for
the metric given by equation \eqref{eq:metricf}. If we
consider the radial path from the emitting object to the 
receiver on Earth, we can fix the angular coordinates to
obtain
\beq
(c+t\dot{c})^2 {\ud t}^2 - a^2 {\ud r}^2 = 0.
\label{eq:geo}
\eeq
where $a(t)$ and $c(t)$ depend on time.
Integrating equation \eqref{eq:geo} one gets:
\begin{align}
&\int_{t_{\textsc{em},1}}^{t_{\textsc{rec},1}}
\frac{c+t\dot{c}}{a(t)} \ud t \,=\,
\int_{r_{\textsc{em},1}}^{r_{\textsc{rec},1}} \ud r
\,=\, r_{\textsc{em},1} - r_{\textsc{rec},1} 
\label{eq:emrec1}\\[2mm]
&\int_{t_{\textsc{em},2}}^{t_{\textsc{rec},2}}
\frac{c+t\dot{c}}{a(t)} \ud t \,=\,
\int_{r_{\textsc{em},2}}^{r_{\textsc{rec},2}} \ud r
\,=\, r_{\textsc{em},2} - r_{\textsc{rec},2}\,.
\label{eq:emrec2}
\end{align}

Assuming that both the emitting object and Earth are at rest
with respect to the cosmological fluid, it follows that the
two radial comoving coordinates for the emission of crest 1
and crest 2 are equal, and the same is true for reception.
Subtracting equation \eqref{eq:emrec2} from equation
\eqref{eq:emrec1} one gets therefore
\beq
\int_{t_{\textsc{em},1}}^{t_{\textsc{rec},1}}
\frac{c+t\dot{c}}{a(t)} \ud t \,-
\int_{t_{\textsc{em},2}}^{t_{\textsc{rec},2}}
\frac{c+t\dot{c}}{a(t)} \ud t \,=\, 0
\eeq
or
\beq
\int_{t_{\textsc{em},1}}^{t_{\textsc{em},2}}
\frac{c+t\dot{c}}{a(t)} \ud t \,-
\int_{t_{\textsc{rec},1}}^{t_{\textsc{rec},2}}
\frac{c+t\dot{c}}{a(t)} \ud t \,=\, 0.
\eeq
Using equations \eqref{eq:w}, \eqref{eq:aditw},
\eqref{eq:cditw} and integrating we get
\beq
\frac{c_0}{a_0 t_0^{N-1}}
\left( t^N_{\textsc{rec},2} - t^N_{\textsc{rec},1} \right) 
\,=\,
\frac{c_0}{a_0 t_0^{N-1}}
\left( t^N_{\textsc{em},2} - t^N_{\textsc{em},1} \right). 
\eeq
Rewriting $t^N_{\textsc{rec},2}$ as
\beq
t^N_{\textsc{rec},2} =
\frac{t^{2N-1}_{\textsc{rec},2}}{t^N_{\textsc{rec},2}}\,
t_{\textsc{rec},2}
\eeq
and the same for $t^N_{\textsc{rec},1}$,
$t^N_{\textsc{em},2}$ and $t^N_{\textsc{em},1}$, and using
again equations \eqref{eq:w}, \eqref{eq:aditw} and
\eqref{eq:cditw}, we obtain
\beq
\left( \frac{ct}{a} \right)_{\textsc{rec},2} -
\left( \frac{ct}{a} \right)_{\textsc{rec},1} 
=\, 
\left( \frac{ct}{a} \right)_{\textsc{em},2} -
\left( \frac{ct}{a} \right)_{\textsc{em},1}\,.
\eeq
Given that
\begin{align}
& c_{\textsc{rec},2} = c_{\textsc{rec},1} \equiv 
c_{\textsc{rec}}; \qquad c_{\textsc{em},2} = c_{\textsc{em},1} 
\equiv  c_{\textsc{em}}\\[2mm]
& a_{\textsc{rec},2} = a_{\textsc{rec},1} \equiv 
a_{\textsc{rec}}; \qquad a_{\textsc{em},2} = a_{\textsc{em},1} 
\equiv  a_{\textsc{em}}
\end{align}
we get
\beq
\left( \frac{c}{a} \right)_{\textsc{rec}} 
(t_{\textsc{rec},2} -t_{\textsc{rec},1})
\,=\,
\left( \frac{c}{a} \right)_{\textsc{em}} 
(t_{\textsc{em},2} -t_{\textsc{em},1}).
\eeq
Using
\beq
c_{\textsc{rec}}(t_{\textsc{rec},2} -t_{\textsc{rec},1}) =
c_{\textsc{rec}}/\nu_{\textsc{rec}} = \lambda_{\textsc{rec}}
\eeq
and the same for emission, we finally obtain
\beq
\frac{\lambda_{\textsc{rec}}}{a_{\textsc{rec}}}  
\,=\,
\frac{\lambda_{\textsc{em}}}{a_{\textsc{em}}}  
\eeq
that is, $\lambda$ and $a$ are proportional as 
we set out to prove.



\begin{thebibliography}{99}
\bibitem{gauss}
We may say that its role is the same as that of the Gaussian
curvature in the theory of 2-D curved surfaces; indeed one
has that $R= -2 \cdot (\text{Gaussian curvature}$) in 3+1
dimensions. It is well known that a surface of constant
curvature has a fixed Gaussian curvature. For instance an
ordinary 2-sphere has a Gaussian curvature given by $1/r^2$
where $r$ is the radius of the sphere. The Ricci scalar for
a sphere is $2/r^2$. If the manifold has more than 2
dimensions or is pseudo-Riemannian, the general argument
holds true.
\end{thebibliography}
\end{document}